\documentclass[12pt]{article}

\newenvironment{pf}{\noindent\textbf{Proof.}~}{\hfill{$\Box$}}

\usepackage{graphicx}
\usepackage{epsfig}
\usepackage{rotating}
\usepackage{amssymb}
\usepackage{amsmath,amsfonts,mathtools}
\usepackage[latin1]{inputenc}
\usepackage{verbatim}
\usepackage[tableposition=top,font=small,labelfont=bf,format=hang]{caption}
\usepackage{booktabs}
\usepackage[T1]{fontenc}
\usepackage{lmodern}
\usepackage{microtype}
\usepackage[hidelinks,pdfusetitle]{hyperref}
\numberwithin{equation}{section}
\newcommand{\coeff}[2]{[#1^{#2}]}
\newtheorem{defi}{Definition}
\newtheorem{lem}{Lemma}
\newtheorem{thm}{Theorem}
\newtheorem{prop}{Proposition}

\newtheorem{rem}{Remark}
\newtheorem{coro}{Corollary}
\newtheorem{ex}{Example}

\newcommand{\F}{\mathbb{F}}
\newcommand{\R}{\mathbb{R}}
\newcommand{\Z}{\mathbb{Z}}

\newenvironment{proof}{\noindent\textbf{Proof.}~}
{\hspace{\stretch{1}}%
\rule{1ex}{1ex}\\}

\DeclareMathOperator{\wt}{wt}

\title{Gaussian mass inequalities for lattices}
\author{Minjia Shi, Xinguang Dai\thanks{Minjia Shi and Xinguang Dai are with the Key Laboratory of Intelligent Computing Signal Processing, Ministry of Education, School of Mathematical Sciences, Anhui
		University, Hefei 230601, China. smjwcl.good@163.com, 	daixg88@163.com}, Patrick Sol\'e\thanks{Institut de Math\'ematiques de Marseille (CNRS, Aix-Marseille University), Marseilles, France. sole@enst.fr}}
\date{}
\begin{document}
\maketitle
\begin{abstract}
The Gaussian mass of a lattice is the evaluation of its theta series at a
real nome and is relevant to lattice coding for the wiretap Gaussian channel.
The ratio of the Gaussian mass of stable lattices to the Gaussian mass of the integer lattice was  conjectured to be $\le 1$ (Regev et al., 2024).
Kominers proved this bound recently for lattices obtained by Construction~A from
binary self-dual codes. Building on Kominers' bound, we establish corresponding inequalities on average
for Construction~$A_4$ lattices obtained from free self-dual quaternary codes and for
Construction~$A_6$ lattices obtained from self-dual $\mathbb Z_6$-codes.
We also prove the strict Gaussian-mass inequality for extremal unimodular
lattices of both Type~II and Type~I. The Type~I result uses the
Rains--Sloane definition of extremality. Its proof combines a universal
theta truncation with an exact inversion of the shadow coefficients, and a
Gaussian inequality for translated lattices.
In addition, we obtain an averaging bound over the genus of Type~II lattices, and
extend the binary Construction~A inequality to Siegel theta series of
arbitrary genus, with further quaternary results under conditions on the
residue code.
\end{abstract}
\ \
{\bf Math Sc. Classification (2020):} 94B05, 11H06, 11H71, 11F11, 11F46

\section{Introduction}
Let \(L\) be a lattice in \(\R^n\). Its theta series is defined by
\[
\Theta_L(q)=\sum_{x\in L}q^{(x,x)},\ 0<q<1,
\]
where \((x,x)\) is the Euclidean inner product.
The Gaussian mass \(G_L(t)\) of \(L\) at \(t\) is the real number obtained upon letting
\(q=\exp(- t)\) in \(\theta_L(q)\). In the sum \(G_L(t)=\sum_{x\in L}e^{-t\|x\|^2}\), each nonzero
vector contributes a weight determined by its length. For large \(t\),
the leading contribution to \(G_L(t)-1\) comes from the shortest
nonzero vectors of \(L\). As \(t\) decreases, vectors of greater
length contribute more substantially, so the Gaussian mass reflects
the distribution of vector lengths beyond the first shell. Applications of this invariant range from geometry of numbers \cite{CKMRV2022}, to computer science \cite{MR}, to cryptography~\cite{Peikert2010}. 

The theoretical study of the Gaussian mass was fundamentally shaped by
the seminal and highly influential work of Regev and
Stephens-Davidowitz~\cite{annals}, who formulated a reverse Minkowski
problem. In earlier work, the same authors~\cite{RSD2017} proved a
striking and elegant inequality that bounds the product of the Gaussian
masses of two translates \(L+u\) and \(L+v\) in terms of the masses of
\(L\), \(L+u+v\), and \(L+u-v\). Special cases of this inequality arise
in the analysis of algorithms for the shortest and closest vector
problems. Building on the framework of discrete Gaussian measures
developed by Micciancio and Regev~\cite{MR}, Regev and
Stephens-Davidowitz, in a particularly insightful line of inquiry, asked
whether every stable lattice \(L\subseteq\R^n\) satisfies
\[
G_L(t)\le G_{\Z^n}(t),\ t>0.
\]
This conjecture asserts that, among all stable lattices of rank \(n\),
the integer lattice \(\Z^n\) maximizes the Gaussian mass at every
scale. Regev and Stephens-Davidowitz proved this inequality for
sufficiently small and sufficiently large \(t\), with thresholds
depending only on \(n\), with remarkable technical precision. Bollauf
and Lin~\cite[Conj.~1]{BL} formulated the corresponding conjecture for
unimodular lattices, a special case of stable lattices.

Building upon these pioneering efforts, S. D. Kominers~\cite{SDK} achieved a significant breakthrough by establishing this bound for lattices \(L\) obtained via Construction A from a binary code. The core of his masterful proof lies in seamlessly bridging the weight enumerator of the code, Jacobi modular forms, and the theta series of the lattice. Beyond these code-based constructions, Kominers~\cite[Theorem~1.1]{SDK2} further made a substantial contribution by resolving the conjecture for general low-dimensional integral unimodular lattices. He proved that every positive-definite integral unimodular lattice \(L\) of rank \(n\le 32\) satisfies \(G_L(t)\le G_{\Z^n}(t)\) for all \(t>0\), and exactly characterized the equality case: it holds for some \(t>0\) if and only if \(L\cong\Z^n\). Furthermore, he pushed this sharp analysis into higher dimensions by proving a strict inequality for every even unimodular lattice of rank \(40\) for all \(t>0\). The technical depth of his approach is once again apparent, as his proof strategically employs root counts and, in the odd case, leverages the nonnegativity of the shadow coefficients to tightly bound the normalized theta series. In Section~\ref{sec:averaging}, we apply his theorem to each Construction~\(A_6\) lattice of rank below \(24\) before averaging over the ternary component.

Beyond its purely mathematical interest, this question is deeply motivated by applications in lattice coding for the Gaussian wiretap channel~\cite{sec} and, more generally, to physical-layer security~\cite{IJLSZ}. For a treatment of semantic security for wiretap channels, see \cite{BTV}. In this context, the {\em secrecy gain} \cite{sec}---defined as the maximum of the ratio \(G_{\Z^n}(t)/G_L(t)\)---is of fundamental significance. The Belfiore-Sol\'e conjecture, which states that the secrecy gain is achieved in the center of the fundamental domain, is proved for most unimodular lattices in dimension \(<32\) \cite{H}. Lattices satisfying a Kominers-like bound inherently possess a secrecy gain \(>1\), thereby providing a mathematically provable safe protection against an eavesdropper~\cite{BL}.

Motivated by the conjecture in \cite{annals} and extending the binary construction of Kominers, this paper establishes the following Gaussian-mass inequalities.
\begin{itemize}
\item Averaged inequalities hold for lattices arising from Construction~\(A_4\) over free self-dual quaternary codes and from Construction~\(A_6\) over self-dual \(\mathbb Z_6\)-codes.
\item Extremal Type~II lattices satisfy the strict pointwise inequality by an argument bearing on the zero location of some special modular forms \cite{JP}. We also bound the genus average of Type~II lattices by the theta series of an orthogonal sum of copies of \(E_8\).
\item Extremal Type~I lattices satisfy the strict pointwise inequality
in every dimension in which they exist, including the exceptional
case of dimension~\(23\) in the Rains--Sloane bound~\cite{RS}.
The proof uses shadow theory~\cite{GSZ,RS} and requires no assumption
of \(s\)-extremality. Extremality determines the initial truncation
of the normalized theta series, which we bound using a positive
Bernstein-basis representation. To bound the remaining terms
uniformly, we combine an exact coefficient inversion in the shadow
expansion with Poisson summation on a translate of the lattice.
\item Using genus-\(g\) weight enumerators and the theta addition formula \cite[Lemma~3.2]{Runge1995}, we extend the binary Construction~A inequality to Siegel theta series of arbitrary genus. We also obtain higher-genus bounds for quaternary codes under conditions on their binary residues.
\end{itemize}

The paper is organized as follows. Section~\ref{sec:preliminaries} collects the preliminaries. Section~\ref{sec:weight-enumerators} relates quaternary weight enumerators to lattice theta series, and Section~\ref{sec:averaging} develops the averaging arguments for Constructions~\(A_4\) and~\(A_6\). Section~\ref{sec:typeII} treats extremal Type~II lattices and their genus averages. Section~\ref{sec:typeI} proves the inequality for extremal Type~I lattices. Finally, Section~\ref{sec:higher-genus} studies the higher-genus inequalities. Section~\ref{conclusion} concludes the article.

{
\section{Preliminaries}
\label{sec:preliminaries}

We recall the basic notation concerning self-dual codes,
Construction~A lattices, Gaussian mass, extremal unimodular
lattices, and higher-genus theta series.

Let $M\geq2$ and write $\Z_M=\Z/M\Z$. A linear code of length $n$ over $\Z_M$ is a $\Z_M$-submodule $C\subseteq\Z_M^n$. Its dual code, with respect to the standard inner product, is
\(
C^\perp
=
\{x\in\Z_M^n:
x\cdot c=0\pmod M
\text{ for every }c\in C\}
\).
The code $C$ is called self-orthogonal if $C\subseteq C^\perp$, and self-dual if $C=C^\perp$. A code is called free of rank $k$ if it is isomorphic, as a
$\Z_M$-module, to $\Z_M^k$.

For $a\in\Z_M$, define its Euclidean weight by
\[
w_E(a)
=
\min\{r^2:r\in\Z,\ r\equiv a\pmod M\},
\]
and, for $c=(c_1,\ldots,c_n)\in\Z_M^n$, put
\(
w_E(c)
=
\sum_{i=1}^n w_E(c_i)
\).
For even $M$, a self-dual code over $\Z_M$ is called Type~II if $w_E(c)\equiv0\pmod{2M}$ for every $c\in C$; otherwise it is called Type~I. In particular,
the divisibility conditions are modulo $8$ over $\Z_4$ and modulo
$12$ over $\Z_6$.

For a linear code $C\subseteq\Z_M^n$, its Construction~A lattice is
\[
A_M(C)
=
\frac{1}{\sqrt M}
\{x\in\Z^n:x\bmod M\in C\},
\]
to simplify some formulas we let $a_M(C)={\sqrt M}.$
If $C$ is self-dual, then $A_M(C)$ is an integral unimodular
lattice. If, in addition, $C$ is Type~II, then $A_M(C)$ is even
unimodular.

For a $\Z_4$-code $C$, its binary residue code is
\(
\operatorname{Res}(C)
=
\{c\bmod2:c\in C\}
\subseteq\mathbb F_2^n.
\)
For each $c\in C$, its residue has Hamming weight
$\wt(c\bmod 2)\equiv c\cdot c\pmod4$. Thus $\operatorname{Res}(C)$
is doubly even whenever $C$ is self-orthogonal.
This binary code will be fixed in the averaging argument for free
self-dual quaternary codes.

For modulus $6$, the Chinese remainder theorem gives $\Z_6 \cong \mathbb F_2\times\mathbb F_3$. Accordingly, a $\Z_6$-code $C$ determines the binary and ternary
components
\(
C_2=C\bmod2
\subseteq\mathbb F_2^n,
\
C_3=C\bmod3
\subseteq\mathbb F_3^n.
\)
Conversely,
\[
C
=
\{c\in\Z_6^n:
c\bmod2\in C_2,\;
c\bmod3\in C_3\}.
\]
Moreover, duality is compatible with the Chinese remainder
decomposition, so that $C$ is self-dual if and only if both $C_2$
and $C_3$ are self-dual. This decomposition will be used in the
averaging argument for Construction $A_6$.

For any finite ensemble $\mathcal E$ and any function $F$ on
$\mathcal E$, we use the notation
\[
\mathbb E_{X\in\mathcal E}[F(X)]
=
\frac1{|\mathcal E|}
\sum_{X\in\mathcal E}F(X).
\]
The two averaging procedures considered later correspond to fixing
the binary component and averaging over the ternary component.

Let $L\subseteq\mathbb R^n$ be a positive-definite lattice. Its theta
series is
\[
\Theta_L(q)
=
\sum_{x\in L}q^{(x,x)},
\
0<q<1.
\]
We use the Jacobi theta function
\(
\vartheta_3(q)
=
\sum_{m\in\Z}q^{m^2}.
\)
Thus
\(
\Theta_{\Z^n}(q)
=
\vartheta_3(q)^n.
\)

Equivalently, putting $q=e^{-\pi t}, \ t>0$, one may write
\[
G_L(t)=\Theta_L(e^{-\pi t})
=
\sum_{x\in L}
e^{-\pi t(x,x)}.
\]
This is the Gaussian mass of $L$ defined in the Introduction.
The Gaussian-mass inequality considered throughout this paper is
\[
\Theta_L(q)
\leq
\vartheta_3(q)^n,
\
0<q<1.
\]
Equivalently,
\[
\sum_{x\in L}e^{-\pi t(x,x)}
\leq
\left(
\sum_{m\in\Z}e^{-\pi tm^2}
\right)^n,
\
t>0.
\]
For lattices obtained by Construction~A from binary self-dual codes,
the corresponding inequality is known (see \cite{SDK}). The purpose
of the averaging results below is to establish analogous inequalities
on average for Constructions $A_4$ and $A_6$.

For later use, we also recall the complete weight enumerator of a
code $C\subseteq\Z_M^n$:
\[
\operatorname{cwe}_C
\bigl(
X_a:a\in\Z_M
\bigr)
=
\sum_{c\in C}
\prod_{i=1}^nX_{c_i}.
\]
For $a\in\Z_M$, define the local theta function
\[
\phi_a^{(M)}(q)
=
\sum_{m\in\Z}
q^{(a+Mm)^2/M}.
\]
By decomposing $A_M(C)$ into cosets of $M\Z^n$, one obtains the
standard Construction~A identity
\[
\Theta_{a_M(C)}(q)
=
\operatorname{cwe}_C
\left(
\phi_a^{(M)}(q):a\in\Z_M
\right).
\]

In the quaternary case, if $n_j(c)$ denotes the number of coordinates
of $c$ equal to $j$, the symmetrized weight enumerator is
\[
\operatorname{swe}_C(X,Y,Z)
=
\sum_{c\in C}
X^{n_0(c)}
Y^{n_1(c)+n_3(c)}
Z^{n_2(c)}.
\]
It is the specialization $\operatorname{cwe}_C(X,Y,Z,Y)$ of the complete weight enumerator. The corresponding relation between
the symmetrized weight enumerator and the theta series of
$A_4(C)$ is the standard code--lattice theta-series correspondence
(see \cite{CS}).

We next recall the lattice terminology needed for the extremal case.
For a positive-definite lattice $L$, its minimum norm is
\[
\min(L)
=
\min\{(x,x):0\neq x\in L\}.
\]
A lattice is unimodular if $L=L^*$, and is even if $(x,x)\in2\Z$ for every $x\in L$.

An integral unimodular lattice is of Type~II if it is even, and
of Type~I if it is odd.

An even unimodular lattice can exist only in dimensions divisible
by $8$. If $L$ is an even unimodular lattice of dimension $n$, then
its minimum norm satisfies the classical bound (see \cite[p.~1489]{NP})
\[
\min(L)
\leq
2\left\lfloor\frac{n}{24}\right\rfloor+2.
\]

\begin{defi}
An even unimodular lattice $L$ of dimension $n$ is called
\emph{extremal} if
\[
\min(L)
=
2\left\lfloor\frac{n}{24}\right\rfloor+2.
\]
\end{defi}

The theta series of an even unimodular lattice of dimension $n$ is
a modular form of weight $n/2$. Consequently, extremality imposes the
maximal possible vanishing of its initial nonconstant Fourier
coefficients. This modular-form description will be used later to
study the Gaussian mass of extremal even unimodular lattices.

We finally introduce the higher-genus setting. For an integer
$g\geq1$, let
\[
\mathbb H_g
=
\{Z\in M_g(\mathbb C):
Z=Z^T,\;
\operatorname{Im}(Z)>0\}
\]
be the Siegel upper half-space of degree $g$.

For $x=(x_1,\ldots,x_g)\in L^g$, let $G(x) = \bigl( (x_i,x_j) \bigr)_{1\leq i,j\leq g}$ denote its Gram matrix.

\begin{defi}
The genus-$g$ theta series of a lattice $L$ is
\[
\Theta_L^{(g)}(Z)
=
\sum_{(x_1,\ldots,x_g)\in L^g}
\exp
\left(
\pi i\,
\operatorname{Tr}
\bigl(G(x)Z\bigr)
\right),
\
Z\in\mathbb H_g.
\]
\end{defi}

For $g=1$ this reduces to the ordinary theta series. If $L$ is even
unimodular of dimension $n$, then $\Theta_L^{(g)}$ is a Siegel modular
form of degree $g$ and weight $n/2$.

For the Gaussian formulation, let $Y=Y^T>0$ be a positive-definite real $g\times g$ matrix and put
\[
G_L^{(g)}(Y)
=
\Theta_L^{(g)}(iY).
\]
Equivalently,
\[
G_L^{(g)}(Y)
=
\sum_{x_1,\ldots,x_g\in L}
\exp
\left(
-\pi
\sum_{a,b=1}^g
Y_{ab}(x_a,x_b)
\right).
\]

Define
\[
\Theta_g(Y)
=
\sum_{m\in\Z^g}
e^{-\pi m^TYm}.
\]
Since $\Z^n$ is an orthogonal direct product of $n$ copies of $\Z$,
we have
\[
G_{\Z^n}^{(g)}(Y)
=
\Theta_g(Y)^n.
\]
Thus the natural genus-$g$ extension of the Gaussian-mass inequality
is
\[
G_L^{(g)}(Y)
\leq
\Theta_g(Y)^n,
\
Y=Y^T>0.
\]
For $g=1$, with $Y=(t)$, this is exactly the ordinary Gaussian-mass
inequality.

Notice also that for the scalar diagonal matrix $Y=tI_g$ the variables separate and
\[
G_L^{(g)}(tI_g)
=
\Theta_L(e^{-\pi t})^g.
\]
Hence the genuinely higher-genus problem arises when $Y$ is allowed
to be an arbitrary positive-definite symmetric matrix.

To relate the higher-genus lattice theta series to codes, let $C\subseteq\Z_M^n$. Its genus-$g$ complete weight enumerator is
\[
\operatorname{cwe}_C^{(g)}
\bigl(
X_a:a\in\Z_M^g
\bigr)
=
\sum_{(c^{(1)},\ldots,c^{(g)})\in C^g}
\prod_{i=1}^n
X_{(c_i^{(1)},\ldots,c_i^{(g)})}.
\]
For $a\in\Z_M^g$ and $Y=Y^T>0$, define
\[
\phi_a^{(M,g)}(Y)
=
\sum_{m\in\Z^g}
\exp
\left(
-\frac{\pi}{M}
(a+Mm)^TY(a+Mm)
\right).
\]
The higher-genus Construction~A identity is
\[
G_{a_M(C)}^{(g)}(Y)
=
\operatorname{cwe}_C^{(g)}
\left(
\phi_a^{(M,g)}(Y):
a\in\Z_M^g
\right).
\]
The identity follows by grouping the integer vectors
$\sqrt{M}x_1,\ldots,\sqrt{M}x_g$ by their residues modulo $M$.
For each fixed tuple in $C^g$, the sum factors over the $n$ coordinates.
In the quaternary case used in Section~\ref{sec:higher-genus},
$\phi_a^{(4,g)}(Y)$ is the theta constant with period matrix $4iY$
and characteristic $\left[\begin{smallmatrix}a/4\\0\end{smallmatrix}\right]$
(see \cite[\S~12]{Mumford1967}).

The ordinary Gaussian-mass problem, its averaged versions for
Constructions $A_4$ and $A_6$, the extremal even-unimodular case,
and the higher-genus formulation above will be considered in the
following sections.}

\section{Theta series of lattices from quaternary codes}
\label{sec:weight-enumerators}

Let $C\subseteq\Z_4^n$ be a linear code. For each $c\in C$, choose its
coordinates in $\{0,1,2,3\}$. The lattice vectors in $A_4(C)$ lying
above $c$ form the coset $c/2+2\Z^n$. The contribution of this coset
to the theta series is the product of the one-dimensional sums
\[
\sum_{u\in\Z^n}q^{\|c/2+2u\|^2}
=\prod_{i=1}^n\phi_{c_i}^{(4)}(q),\ 0<q<1.
\]
Here $\phi_a^{(4)}(q)$ is the sum over $a/2+2\Z$ defined in
Section~\ref{sec:preliminaries}. To evaluate these four sums, write
\[
\vartheta_2(q)=\sum_{m\in\Z}q^{(m+1/2)^2}.
\]
The classes $a=0$ and $a=2$ give the even and odd integers, respectively,
so $\phi_0^{(4)}(q)=\vartheta_3(q^4)$ and
$\phi_2^{(4)}(q)=\vartheta_2(q^4)$. The coordinate sets for $a=1$ and
$a=3$ are negatives of one another, and their disjoint union is
$\Z+\tfrac12$. Their theta sums are therefore equal, each contributing
$\vartheta_2(q)/2$. Summing over the codewords now gives
\[
\Theta_{A_4(C)}(q)
=\operatorname{swe}_C\!\left(\vartheta_3(q^4),\frac{\vartheta_2(q)}2,\vartheta_2(q^4)\right).
\]
This is the specialization at $z=0$ of the theta identity of Choie and
Sol\'e~\cite[Theorem~4.1]{CS}. Bollauf, Lin, and Ytrehus give the
displayed formula in \cite[Corollary~25]{BLY} and its one-variable
normalization in \cite[Theorem~28]{BLY}.

\begin{lem}\label{lem:main-reduction}
Let $C\subseteq\Z_4^n$ be self-dual. For $0<r<1$, define
\[
H_C(r)=\operatorname{swe}_C\!\left(1+r,(1-r^4)^{1/4},1-r\right).
\]
Put $r(q)=\vartheta_4(q)/\vartheta_3(q)$, where
$\vartheta_4(q)=\sum_{m\in\Z}(-1)^m q^{m^2}$.
Then $r(q)$ takes every value in $(0,1)$ as $q$ ranges over $(0,1)$,
and
\begin{equation}\label{eq:legacy-4-1}
\frac{\Theta_{A_4(C)}(q)}{\Theta_{\Z^n}(q)}=2^{-n}H_C(r(q)).
\end{equation}
\end{lem}

\begin{proof}
Write $r=r(q)$. Separating the even and odd indices in the two defining
series gives
\[
\begin{aligned}
\vartheta_3(q)&=\vartheta_3(q^4)+\vartheta_2(q^4),\\
\vartheta_4(q)&=\vartheta_3(q^4)-\vartheta_2(q^4).
\end{aligned}
\]
Adding and subtracting these identities yields
\[
\frac{\vartheta_3(q^4)}{\vartheta_3(q)}=\frac{1+r}{2},\
\frac{\vartheta_2(q^4)}{\vartheta_3(q)}=\frac{1-r}{2}.
\]
For the odd residue classes, Jacobi's identity
$\vartheta_2(q)^4=\vartheta_3(q)^4-\vartheta_4(q)^4$ gives
$\vartheta_2(q)/(2\vartheta_3(q))=(1-r^4)^{1/4}/2$, with the
positive fourth root. Each monomial of $\operatorname{swe}_C$ has
degree $n$. To divide the theta series by $\Theta_{\Z^n}(q)=\vartheta_3(q)^n$,
we may therefore divide each argument of the weight enumerator by
$\vartheta_3(q)$. Pulling out the common factor $1/2$ from the three
arguments gives $2^{-n}$ in \eqref{eq:legacy-4-1}.

To determine the range of $r$, write $q=e^{-\pi t}$. Jacobi's
transformation~\cite[\S~20.7(viii)]{NIST} gives
\[
r(e^{-\pi t})=\frac{\vartheta_2(e^{-\pi/t})}{\vartheta_3(e^{-\pi/t})}>0.
\]
Also $\vartheta_3(q)-\vartheta_4(q)=2\vartheta_2(q^4)>0$, so $r(q)<1$.
As $q\to0^+$, both $\vartheta_3(q)$ and $\vartheta_4(q)$ tend to $1$.
As $q\to1^-$, we have $t\to0^+$ and $e^{-\pi/t}\to0$, so the
transformed ratio tends to $0$. Continuity now gives the stated range.
\end{proof}

For a self-dual code, $H_C$ extends to a polynomial. Indeed, the
exponent of the middle argument of $\operatorname{swe}_C$ in the term
indexed by $c$ is $n_1(c)+n_3(c)=\wt(c\bmod 2)$. This is a multiple
of four by the residue-code property in Section~\ref{sec:preliminaries},
so the fourth root disappears in each term. At $r=0$ every codeword
contributes $1$, giving $H_C(0)=|C|=2^n$ by self-duality. At $r=1$
only the zero codeword contributes, giving $H_C(1)=2^n$.

By \eqref{eq:legacy-4-1} and the range of $r(q)$, the Gaussian-mass
bound for $A_4(C)$ at every $q\in(0,1)$ is equivalent to
$H_C(r)\le 2^n$ for all $0<r<1$. The same reduction applies to
an average over any nonempty finite family $\mathcal E$ of self-dual
codes of length $n$. Since $\Theta_{\Z^n}(q)$ is independent of $C$, we have
\[
\frac{\mathbb E_{C\in\mathcal E}\bigl[\Theta_{A_4(C)}(q)\bigr]}{\Theta_{\Z^n}(q)}
=2^{-n}\mathbb E_{C\in\mathcal E}\bigl[H_C(r(q))\bigr].
\]
In Section~\ref{sec:averaging} the quaternary codes have length $2k$
and the binary residue matrix is fixed. For this family the required
upper bound on the average of $H_C(r)$ is $2^{2k}=4^k$.

\section{\texorpdfstring{Averaging bound for constructions $A_4$ and $A_6$}{Averaging bound for constructions A4 and A6}}
\label{sec:averaging}
Let $C$ be a free self-dual $\mathbb Z_4$-code of length $n=2k$. After a permutation of coordinates, write its generator matrix $G$ in systematic form as
\[
G=[I_k\mid A], \  AA^T\equiv -I_k \pmod 4.
\]
For $0<r<1$, set
\[
f_r(0)=1+r,\ f_r(1)=f_r(3)=(1-r^4)^{1/4},\ f_r(2)=1-r,
\]
and put $F_r(u)=\prod_{i=1}^k f_r(u_i)$ for $u\in\Z_4^k$.
The codewords of $C$ are precisely the pairs $(a,aA)$ with $a\in\Z_4^k$.
Thus in this representation
\[
H_A(r):=H_C(r)=\sum_{a\in\Z_4^k}F_r(a)F_r(aA).
\]
Since $n=2k$, equation~\eqref{eq:legacy-4-1} gives
\[
\frac{\Theta_{A_4(C)}(q)}{\Theta_{\Z^{2k}}(q)}=4^{-k}H_A(r(q)).
\]
Since $A$ is invertible, the Kominers-type identity is
\[
H_A(r)=\Lambda(r)^k-
\frac12\sum_{a\in\mathbb Z_4^k}
\bigl(F_r(a)-F_r(aA)\bigr)^2,
\]
where
\[
\Lambda(r)=2\left(1+r^2+\sqrt{1-r^4}\right).
\]
Thus the desired inequality $H_A(r)\le 4^k$ is equivalent to
\begin{equation}
\frac12\sum_{a\in\mathbb Z_4^k}
\bigl(F_r(a)-F_r(aA)\bigr)^2
\ge \Lambda(r)^k-4^k.
\label{eq:legacy-1}
\end{equation}
The question considered here is whether the left-hand side of \eqref{eq:legacy-1} can be controlled after averaging over the matrices $A$ satisfying $AA^T\equiv-I_k\pmod4$.
Put $B=A\bmod2$. Then $BB^T=I_k$ over $\mathbb F_2$. The associated binary code
\[
D_B=\{(x,xB):x\in\mathbb F_2^k\}
\]
is self-dual. It is also doubly even, since it is the residue code of
the self-orthogonal code $C$.

Fix one lift $A_0$ of $B$ satisfying $A_0A_0^T\equiv-I_k\pmod4$. Every other lift with the same residue matrix can be written uniquely as
\begin{equation}
A_H=A_0+2HB,
\label{eq:legacy-2}
\end{equation}
where $H$ runs through the symmetric matrices in $M_k(\mathbb F_2)$.

Indeed, write $A=A_0+2E$. Comparing the two congruences $AA^T\equiv A_0A_0^T\equiv-I_k\pmod4$ gives $EB^T+BE^T=0$ over $\mathbb F_2$. If $H=EB^T$, then $H+H^T=0$, which in characteristic $2$ means $H=H^T$. Since $BB^T=I_k$, one has $E=HB$, giving \eqref{eq:legacy-2}.

Thus it is enough first to average over the symmetric binary matrix $H$, while keeping $B$ fixed.

\subsection{\texorpdfstring{Exact average over all lifts of a fixed $B$}{Exact average over all lifts of a fixed B}}

Set $\alpha=1+r$, $\gamma=1-r$, and
$\eta=(1-r^4)^{1/4}$.
Write every $a\in\mathbb Z_4^k$ uniquely as
\[
a=x+2v,\  x,v\in\mathbb F_2^k.
\]

Assume first that $x\ne0$. From \eqref{eq:legacy-2},
\begin{equation}
aA_H\equiv aA_0+2xHB\pmod4.
\label{eq:legacy-3}
\end{equation}
As $H$ runs uniformly over all symmetric binary matrices, the vector $xH$ runs uniformly over $\mathbb F_2^k$. Hence $xHB$ is also uniform, because $B$ is invertible. Therefore $aA_H$ is uniformly distributed among all vectors in $\mathbb Z_4^k$ whose residue modulo $2$ is $xB$.

For a coordinate with binary residue $0$, the two possible $\mathbb Z_4$ values are $0$ and $2$, and their average contribution to $F_r$ is $\frac{\alpha+\gamma}{2}=1$. For a coordinate with binary residue $1$, the two possible values are $1$ and $3$, and both contribute $\eta$. Consequently,
\begin{equation}
\mathbb E_{H}\,F_r(aA_H)=\eta^{\operatorname{wt}(xB)},\  x\ne0.
\label{eq:legacy-4}
\end{equation}
On the other hand,
\begin{equation}
\sum_{v\in\mathbb F_2^k}F_r(x+2v)
=2^k\eta^{\operatorname{wt}(x)}.
\label{eq:legacy-5}
\end{equation}
Combining \eqref{eq:legacy-4} and \eqref{eq:legacy-5}, the total contribution from $x\ne0$ is
\begin{equation}
2^k\sum_{x\ne0}
\eta^{\operatorname{wt}(x)+\operatorname{wt}(xB)}
=2^k\bigl(W_{D_B}(1,\eta)-1\bigr),
\label{eq:legacy-6}
\end{equation}
where $W_{D_B}$ denotes the binary weight enumerator of $D_B$.

Now consider $x=0$. Then $a=2v$ and $2vA_H\equiv2vB\pmod4$, so this part is independent of the lift. Its contribution is
\begin{equation}
\sum_{v\in\mathbb F_2^k}F_r(2v)F_r(2vB)
=W_{D_B}(\alpha,\gamma).
\label{eq:legacy-7}
\end{equation}
Since $D_B$ is binary self-dual, the MacWilliams identity gives
\begin{equation}
\begin{aligned}
W_{D_B}(\alpha,\gamma)
=2^{-k}W_{D_B}(\alpha+\gamma,\alpha-\gamma)
=2^kW_{D_B}(1,r).
\end{aligned}
\label{eq:legacy-8}
\end{equation}
Equations \eqref{eq:legacy-6}--\eqref{eq:legacy-8} yield the exact average formula
\begin{equation}
\mathbb E_{A\mid B}H_A(r)
=2^k\left(W_{D_B}(1,r)+W_{D_B}(1,\eta)-1\right).
\label{eq:legacy-9}
\end{equation}

\subsection{ Convexity argument}
Because $D_B$ is doubly even, there is a polynomial with nonnegative coefficients
\[
P_B(t)=\sum_{j\ge0}A_{4j}t^j
\]
such that $W_{D_B}(1,z)=P_B(z^4)$. Put $t=r^4$. Since $\eta^4=1-r^4=1-t$, formula \eqref{eq:legacy-9} becomes
\begin{equation}
\mathbb E_{A\mid B}H_A(r)
=2^k\left(P_B(t)+P_B(1-t)-1\right).
\label{eq:legacy-10}
\end{equation}
The coefficients of $P_B$ are nonnegative, so $P_B$ is convex on $[0,1]$. Hence $Q_B(t)=P_B(t)+P_B(1-t)$ is also convex. It is symmetric about $t=1/2$, and its endpoint values are equal. Therefore its maximum on $[0,1]$ is attained at the endpoints. Since $P_B(0)=1$ and $P_B(1)=|D_B|=2^k$, we obtain
\begin{equation}
P_B(t)+P_B(1-t)\le1+2^k.
\label{eq:legacy-11}
\end{equation}
Substituting \eqref{eq:legacy-11} into \eqref{eq:legacy-10} gives
\begin{equation}
\mathbb E_{A\mid B}H_A(r)\le4^k.
\label{eq:legacy-12}
\end{equation}
Equivalently,
\begin{equation}
\mathbb E_{A\mid B}
\left[
\frac12\sum_{a\in\mathbb Z_4^k}
\bigl(F_r(a)-F_r(aA)\bigr)^2
\right]
\ge\Lambda(r)^k-4^k.
\label{eq:legacy-13}
\end{equation}
Thus the desired lower bound in \eqref{eq:legacy-1} is valid after averaging over all $\mathbb Z_4$ lifts of every fixed binary residue matrix $B$.

The amount by which the average exceeds the required lower bound can also be written explicitly. From \eqref{eq:legacy-10},
\[
4^k-\mathbb E_{A\mid B}H_A(r)
=2^k\left(1+2^k-P_B(t)-P_B(1-t)\right).
\]
Using $P_B(t)=\sum_jA_{4j}t^j$ gives
\begin{equation}
4^k-\mathbb E_{A\mid B}H_A(r)
=2^k\sum_{j\ge2}A_{4j}
\left(1-t^j-(1-t)^j\right),
\label{eq:legacy-14}
\end{equation}
and every term on the right-hand side is nonnegative for $0\le t\le1$.\\ \ \

\begin{ex}
\normalfont
For $k=4$, the binary doubly-even self-dual code has weight enumerator
\[
W(1,z)=1+14z^4+z^8,\ \text{so }P(t)=1+14t+t^2.
\]
Formula \eqref{eq:legacy-10} gives
\[
\mathbb E_{A\mid B}H_A(r)
=256-32t(1-t),\  t=r^4.
\]
Thus
\[
\begin{aligned}
\mathbb E_{A\mid B}
\left[\frac12\sum_a\bigl(F_r(a)-F_r(aA)\bigr)^2\right]
=\Lambda(r)^4-256+32r^4(1-r^4),
\end{aligned}
\]
which is strictly stronger than the required averaged inequality for $0<r<1$.
\end{ex}

{

\subsection{\texorpdfstring{The exact fixed-binary average for Construction $A_6$}{The exact fixed-binary average for Construction A6}}
\label{subsec:z6-exact-average}

We now turn to Construction $A_6$. In contrast with the
$\mathbb Z_4$ case, where one fixes the binary residue code and
averages over its self-dual lifts, the Chinese remainder decomposition $\mathbb Z_6\cong\mathbb F_2\times\mathbb F_3$ leads to a different natural averaging ensemble.

Let $n=2k=4s>0$ and fix a binary self-dual code $D\subseteq\mathbb F_2^n$. Let
\[
\mathcal S_3(n)
=
\{T\subseteq\mathbb F_3^n:T=T^\perp\},
\]
equipped with the uniform distribution on its actual subspaces.
For $T\in\mathcal S_3(n)$, let
\[
C(D,T)
=
\{c\in\mathbb Z_6^n:
c\bmod2\in D,\;
c\bmod3\in T\},
\]
and put $K=A_2(D)$. Thus the binary component is fixed, while the ternary self-dual
component is averaged.

The basic incidence probability for the ternary ensemble is the
following.

\begin{lem}
\label{lem:z6-incidence}
Let $c\in\mathbb F_3^{2k}$ and let
$T\in\mathcal S_3(2k)$ be chosen uniformly. Then
\begin{equation}
\mathbb P(c\in T)
=
\begin{cases}
1,
& c=0,
\\[1mm]
\dfrac{1}{3^{k-1}+1},
& c\neq0\text{ and }c\cdot c=0,
\\[3mm]
0,
& c\cdot c\neq0.
\end{cases}
\label{eq:z6-incidence}
\end{equation}
\end{lem}

\begin{proof}
We refer to chapter 1 of \cite{K} for working knowledge of quadratic forms over finite fields.
Since $n=2k$ is divisible by $4$, the standard quadratic space
$\mathbb F_3^{2k}$ is split. In a hyperbolic basis, its isotropic
vectors may be represented by pairs $(x,y)\in\mathbb F_3^k\times\mathbb F_3^k$ satisfying $x\cdot y=0$. For $x=0$ there are $3^k$ choices for $y$, whereas for each nonzero
$x$ there are $3^{k-1}$ choices. Hence the number of nonzero isotropic
vectors is
\[
3^k+(3^k-1)3^{k-1}-1
=
(3^k-1)(3^{k-1}+1).
\]

The orthogonal group acts transitively on the nonzero isotropic
vectors and permutes the members of $\mathcal S_3(2k)$. Hence the
probability that a fixed nonzero isotropic vector belongs to a
uniformly chosen $T$ is constant.

Every $T\in\mathcal S_3(2k)$ contains $3^k-1$ nonzero vectors.
Double counting incidences therefore gives
\[
\mathbb P(c\in T)
=
\frac{3^k-1}
{(3^k-1)(3^{k-1}+1)}
=
\frac{1}{3^{k-1}+1}.
\]
A nonisotropic vector cannot belong to a self-orthogonal subspace.
\end{proof}

Define the mod-$3$ norm filter
\begin{equation}
P_3(K;t)
=
\sum_{\substack{v\in K\\
\lVert v\rVert^2\equiv0\pmod3}}
e^{-\pi t\lVert v\rVert^2/3}.
\label{eq:z6-P3}
\end{equation}
Since $K$ is integral, the congruence condition in
\eqref{eq:z6-P3} is well-defined.

\begin{thm}
\label{thm:z6-exact-average}
For every $t>0$,
\begin{equation}
{
\mathbb E_{T\in\mathcal S_3(n)}
\Theta_{A_6(C(D,T))}(t)
=
\frac{
3^{k-1}\Theta_K(3t)+P_3(K;t)
}{
3^{k-1}+1
}.
}
\label{eq:z6-exact-average}
\end{equation}
\end{thm}

\begin{proof}
Write $v=\frac{u}{\sqrt2}\in K$, where $u\in\mathbb Z^n, \ u\bmod2\in D$. The corresponding potential vector in the modulus-$6$
Construction~A lattice is $\frac{u}{\sqrt6} = \frac{v}{\sqrt3}$, and hence its Gaussian contribution is $e^{-\pi t\lVert v\rVert^2/3}$.

The ternary residue $u\bmod3$ can occur in a ternary self-dual code
only when it is isotropic. Since $\lVert u\rVert^2=2\lVert v\rVert^2$, and $2$ is invertible modulo $3$, this is equivalent to $\lVert v\rVert^2\equiv0\pmod3$.

Put $\rho=\frac{1}{3^{k-1}+1}$. By Lemma~\ref{lem:z6-incidence}, every nonzero isotropic ternary
residue occurs with probability $\rho$.

The zero ternary residue means $u=3w$ for some $w\in\mathbb Z^n$. Since $3\equiv1\pmod2$, the condition $u\bmod2\in D$ is equivalent to $w\bmod2\in D$. Hence $\frac{w}{\sqrt2}\in K$. Thus the vectors $v$ with zero ternary residue are exactly the
vectors of $3K$. Their contribution is
\[
\sum_{x\in K}
e^{-\pi t\lVert3x\rVert^2/3}
=
\Theta_K(3t).
\]

Consequently,
\begin{align*}
\mathbb E_T
\Theta_{A_6(C(D,T))}(t)
&=
\Theta_K(3t)
+
\rho
\left(
P_3(K;t)-\Theta_K(3t)
\right)
\\
&=
(1-\rho)\Theta_K(3t)
+
\rho P_3(K;t).
\end{align*}
Since $1-\rho = \frac{3^{k-1}}{3^{k-1}+1}$, equation \eqref{eq:z6-exact-average} follows.
\end{proof}

Thus, after averaging over the ternary self-dual component, all
dependence on $T$ disappears. The resulting expression depends only
on the fixed binary Construction~A lattice $K=A_2(D)$ and its
mod-$3$ norm filter $P_3(K;t)$. This is the modulus-$6$ analogue of
the fixed-residue averaging formula obtained above for
Construction $A_4$.


\subsection{\texorpdfstring{The averaged Gaussian bound for Construction $A_6$}{The averaged Gaussian bound for Construction A6}}
\label{subsec:z6-gaussian-bound}

We next use the exact formula \eqref{eq:z6-exact-average} to derive
the Gaussian-mass bound. The main input is the Gaussian inequality for
binary self-dual Construction~A lattices \cite{SDK}:
\begin{equation}
\Theta_K(u)
\leq
\vartheta_3(u)^n,
\
u>0.
\label{eq:z6-binary-bound}
\end{equation}
For convenience, write
\(
w(t)=\vartheta_3(t),
\
\phi(t)=-\frac{w'(t)}{w(t)}.
\)
We first record the elementary estimates required below.

\begin{lem}
\label{lem:z6-theta-estimates}
One has
\(
w(3)<1.0002,
\
w(30/11)<1.0004,
\
w(11/10)>1.062,
\)
and for $t\geq\frac32$, we have $\phi(t)<1/16$. In addition,
\[
\frac{w(2/3)}{\sqrt3}<\frac34.
\]
Moreover,
\[
(1.062)^{24}
>
(1.0002)^{24}
+
3(1.0004)^{24},
\]
\[
2\left(\frac{24}{25}\right)^{24}<\frac45,\
\text{and }\frac{3\cdot540}{48}\left(\frac34\right)^{24}<\frac1{20}.
\]
\end{lem}

\begin{proof}
Let $q=e^{-\pi t}$. Since $j^2\geq1+3(j-1)$ for every integer $j\geq1$, we have
\[
1+2q
\leq
w(t)
\leq
1+\frac{2q}{1-q^3}.
\]
Similarly,
\[
\phi(t)
\leq
2\pi q
\frac{1+q^3}{(1-q^3)^3}.
\]

Using
\[
e^{-3\pi}<\frac1{12000},
\
e^{-30\pi/11}<\frac1{5200},
\
e^{-11\pi/10}>\frac{31}{1000},
\]
gives the first three estimates.

For $t\geq3/2$, one has $q<\frac1{110}$. Together with $2\pi<44/7$, this yields
\[
\phi(t)
<
\frac{44}{770}
\frac{1+110^{-3}}{(1-110^{-3})^3}
<
\frac1{16}.
\]

The theta transformation gives
\[
\frac{w(2/3)}{\sqrt3}
=
\frac{w(3/2)}{\sqrt2}.
\]
The preceding geometric estimate gives $w(3/2)<1.019$, and therefore
\[
\frac{w(2/3)}{\sqrt3}
<
\frac{1.019}{\sqrt2}
<
\frac34.
\]
The remaining assertions are direct rational comparisons.
\end{proof}
\begin{prop}
\label{prop:z6-analytic-bound}
Let $K$ be an integral unimodular lattice of rank $n=2k\geq24$ satisfying \eqref{eq:z6-binary-bound}. Then
\begin{equation}
\frac{
3^{k-1}\Theta_K(3t)+P_3(K;t)
}{
3^{k-1}+1
}
<
w(t)^n,
\
t\geq1.
\label{eq:z6-analytic-bound}
\end{equation}
\end{prop}

\begin{proof}
Put $a=3^{k-1}$ and
\[
\mathcal B_K(t)
=
\frac{
a\Theta_K(3t)+P_3(K;t)
}{
a+1
}.
\]
First suppose $1\leq t\leq2$. Dropping the norm restriction in $P_3(K;t)$ and using
\eqref{eq:z6-binary-bound} give
\[
P_3(K;t)
\leq
\Theta_K(t/3)
\leq
w(t/3)^n.
\]
By the theta transformation,
\[
w(t/3)
=
\sqrt{\frac3t}\,w(3/t).
\]
Since $a=3^{k-1}=3^{n/2-1}$, it follows that
\[
\mathcal B_K(t)
\leq
\frac{a}{a+1}
\left(
w(3t)^n
+
3t^{-n/2}w(3/t)^n
\right).
\]
It is therefore sufficient to prove
\begin{equation}
w(3t)^n
+
3t^{-n/2}w(3/t)^n
<
w(t)^n,
\
1\leq t\leq2.
\label{eq:z6-central-target}
\end{equation}
Suppose first that $1\leq t\leq\frac{11}{10}$. By monotonicity of $w$,
\[
w(3t)^n
+
3t^{-n/2}w(3/t)^n
<
(1.0002)^n
+
3(1.0004)^n,
\]
whereas $w(t)^n>(1.062)^n$. For every $n\geq24$, Lemma~\ref{lem:z6-theta-estimates} proves
\eqref{eq:z6-central-target}.

Now suppose $\frac{11}{10}\leq t\leq2$ and put $q=e^{-\pi t}$. Since $w(t)>w(3t)\geq1$,
\[
w(t)^n-w(3t)^n
\geq
n\bigl(w(t)-w(3t)\bigr)
\geq
2nq(1-q^2).
\]
Define
\[
F_n(t)
=
\frac{3}{2n}
e^{\pi t}
\left(
\frac{w(3/t)}{\sqrt t}
\right)^n.
\]
Then
\[
\frac{d}{dt}\log F_n(t)
=
\pi-\frac{n}{2t}
+
\frac{3n}{t^2}\phi(3/t).
\]
By Lemma~\ref{lem:z6-theta-estimates}, $\phi(3/t)<\frac1{16}$, and hence
\[
\begin{aligned}
\frac{d}{dt}\log F_n(t)
<\pi+n\left(-\frac1{2t}+\frac3{16t^2}\right)
\leq\pi-\frac{13n}{64}<0.
\end{aligned}
\]
Thus $F_n$ is decreasing.
Furthermore,
\[
\frac{w(30/11)}{\sqrt{11/10}}
<
\frac{24}{25},
\]
so
\[
\begin{aligned}
F_n(t)\leq F_n(11/10)
<\frac{48}{n}\left(\frac{24}{25}\right)^n
\leq2\left(\frac{24}{25}\right)^{24}<\frac45.
\end{aligned}
\]
On the other hand,
\[
1-q^2
\geq
1-e^{-11\pi/5}
>
\frac{99}{100}.
\]
Hence $F_n(t)<1-q^2$, and \eqref{eq:z6-central-target} follows.

It remains to consider $t\geq2$. Put $G_n(t)=w(t)^n-1$. The scalar theta series satisfies
\[
w(3t)-1
\leq
e^{-2\pi t}\bigl(w(t)-1\bigr).
\]
Using
\[
(1+cx)^n-1
\leq
c\bigl((1+x)^n-1\bigr)
\]
for $x\geq0$ and $0\leq c\leq1$, we obtain
\begin{equation}
\Theta_K(3t)-1
\leq
w(3t)^n-1
\leq
e^{-2\pi t}G_n(t).
\label{eq:z6-tail-theta}
\end{equation}

Since $K$ is integral, every nonzero vector counted by $P_3(K;t)$
has squared norm at least $3$. Therefore
\begin{equation}
P_3(K;t)-1
\leq
e^{-\pi(t-2)}
\bigl(P_3(K;2)-1\bigr)
\leq
e^{-\pi(t-2)}w(2/3)^n.
\label{eq:z6-tail-filter}
\end{equation}

Moreover, $G_n(t)\geq2ne^{-\pi t}$. Combining \eqref{eq:z6-tail-theta} and
\eqref{eq:z6-tail-filter} gives
\[
\mathcal B_K(t)-1
\leq
\left[
e^{-2\pi t}
+
\frac{3e^{2\pi}}{2n}
\left(
\frac{w(2/3)}{\sqrt3}
\right)^n
\right]
G_n(t).
\]
Using
\[
e^{-4\pi}<\frac1{100000},
\
e^{2\pi}<540,
\
\frac{w(2/3)}{\sqrt3}<\frac34,
\]
we obtain
\[
\mathcal B_K(t)-1
<
\left[
\frac1{100000}
+
\frac{3\cdot540}{2n}
\left(\frac34\right)^n
\right]
G_n(t)
<
G_n(t).
\]
Hence $\mathcal B_K(t)<w(t)^n$ for every $t\geq2$, completing the proof.
\end{proof}

\begin{thm}
\label{thm:z6-average-gaussian}
Let $D\subseteq\mathbb F_2^n, \ n=4s>0$, be a binary self-dual code. Then
\begin{equation}
{
\mathbb E_{T\in\mathcal S_3(n)}
\Theta_{A_6(C(D,T))}(t)
\leq
\vartheta_3(t)^n
}
\
\text{for every }t>0.
\label{eq:z6-average-gaussian}
\end{equation}
The inequality is strict if $n\geq24$. If $D$ is Type~II, then it is
strict for every positive admissible length.
\end{thm}

\begin{proof}
Let $K=A_2(D)$. By the Gaussian-mass inequality for binary self-dual Construction~A
lattices \cite{SDK},
\[
\Theta_K(u)
\leq
\vartheta_3(u)^n
\]
for every $u>0$.

If $n\geq24$, Theorem~\ref{thm:z6-exact-average} and
Proposition~\ref{prop:z6-analytic-bound} give
\[
\mathbb E_T
\Theta_{A_6(C(D,T))}(t)
<
\vartheta_3(t)^n
\]
for every $t\geq1$.

Each lattice $A_6(C(D,T))$ is integral unimodular. Therefore
Poisson summation gives
\[
\begin{aligned}
\Theta_{A_6(C(D,T))}(t)=t^{-n/2}\Theta_{A_6(C(D,T))}(1/t),
\text{while }\vartheta_3(t)^n=t^{-n/2}\vartheta_3(1/t)^n.
\end{aligned}
\]
Averaging the first identity extends the strict inequality to
$0<t<1$.

It remains to consider the admissible lengths
\mbox{$n\in\{4,8,12,16,20\}$} below $24$. In these dimensions, the Gaussian-mass upper bound for integral
unimodular lattices follows from
\cite[Theorem~1.1]{SDK2}. Therefore
\[
\Theta_{A_6(C(D,T))}(t)
\leq
\vartheta_3(t)^n
\]
for every $T$, and hence the same inequality holds after averaging.

If $D$ is Type~II, then every $C(D,T)$ is Type~II and hence every
$A_6(C(D,T))$ is even unimodular. The remaining positive admissible
lengths below $24$ are $8$ and $16$. In these dimensions such a lattice
cannot be isometric to $\mathbb Z^n$, and the strictness statement in
\cite[Theorem~1.1]{SDK2} therefore yields
\[
\Theta_{A_6(C(D,T))}(t)
<
\vartheta_3(t)^n.
\]
This proves the final assertion.
\end{proof}

When $D$ is Type~II, the exact average admits a natural
Hecke-theoretic interpretation. Let
\[
f_K(\tau)
=
\sum_{v\in K}
e^{\pi i\tau\lVert v\rVert^2},
\
\operatorname{Im}\tau>0.
\]
Since $K$ is even unimodular, write $\lVert v\rVert^2=2m_v, \ m_v\in\mathbb Z$. The root-of-unity filter gives
\begin{equation}
P_3(K;t)
=
\frac13
\sum_{j=0}^{2}
f_K\left(\frac{it+j}{3}\right).
\label{eq:z6-root-filter}
\end{equation}

Define
\begin{equation}
(T_3f)(\tau)
=
3^{k-1}f(3\tau)
+
\frac13
\sum_{j=0}^{2}
f\left(\frac{\tau+j}{3}\right).
\label{eq:z6-Hecke}
\end{equation}

\begin{coro}
\label{coro:z6-hecke}
If $D$ is Type~II, then
\begin{equation}
{
\mathbb E_{T\in\mathcal S_3(n)}
\Theta_{A_6(C(D,T))}(t)
=
\frac{
(T_3f_K)(it)
}{
3^{k-1}+1
}.
}
\label{eq:z6-hecke-average}
\end{equation}
Consequently,
\[
(T_3f_K)(it)
<
(3^{k-1}+1)\vartheta_3(t)^n
\]
for every $t>0$.
\end{coro}

\begin{proof}
Equation \eqref{eq:z6-hecke-average} follows by substituting
\eqref{eq:z6-root-filter} and \eqref{eq:z6-Hecke} into
\eqref{eq:z6-exact-average}. The strict inequality follows from
Theorem~\ref{thm:z6-average-gaussian}.
\end{proof}

\begin{rem}
\label{rem:z6-quantifiers}
Theorem~\ref{thm:z6-average-gaussian} is an averaged statement.
For every fixed $t>0$, it implies that there exists at least one
ternary self-dual component $T$ such that
\[
\Theta_{A_6(C(D,T))}(t)
\leq
\vartheta_3(t)^n.
\]
It does not imply that every ternary component satisfies the Gaussian
bound, nor does it imply the existence of a single fixed $T$ for which
the inequality holds simultaneously for all $t>0$.
\end{rem}
}

\section{Type II lattices}
\label{sec:typeII}
\subsection{Extremal Type II lattices}
\label{subsec:typeII-extremal}
Recall that an even unimodular lattice $L$ of rank $n$ is called
extremal if
\[
\min(L)=2\left\lfloor\frac{n}{24}\right\rfloor+2.
\]
{Even unimodular lattices can exist only when $n$ is a multiple of
$8$ \cite[pp.~191--192]{SPLAG}.}

\begin{thm}
Let $L$ be  an extremal even unimodular lattice of dimension $n$. Its Gaussian mass satisfies
\[
 {\Theta_{L}(q)<\vartheta_3(q)^n,\ 0<q<1.}
\]
\end{thm}

\begin{proof}
Write $n=8m$ and put $\nu=\left\lfloor\frac{m}{3}\right\rfloor=\left\lfloor\frac{n}{24}\right\rfloor$. For $\tau$ in the upper half-plane, write $Q=e^{2\pi i\tau}$ and use the modular-form normalization
\[
 \widehat\Theta_L(\tau)=\sum_{x\in L}Q^{\|x\|^2/2}.
\]
Since $L$ is extremal, $\min L=2(\nu+1)$, and therefore
\[
 \widehat\Theta_L(\tau)=1+O(Q^{\nu+1}).
\]
The theta series $\widehat\Theta_L$ is a modular form of weight $4m$ for $SL_2(\Z)$. Since $\dim M_{4m}(SL_2(\Z))=\nu+1$, the conditions on the first $\nu$ nonconstant Fourier coefficients determine it uniquely. Thus $\widehat\Theta_L$ is the so-called gap modular form of weight $4m$. It can be written in the form
\[
 \widehat\Theta_L(\tau)=E_4(\tau)^mP_m\!\left(\frac{\Delta(\tau)}{E_4(\tau)^3}\right),
\]
where $P_m(X)$ is a real polynomial of degree at most $\nu$ and $P_m(0)=1$. By the theorem of Duke and Jenkins (see \cite[Theorem~1.1]{JP}),
every zero of $\widehat\Theta_L$ in the standard fundamental domain
lies on its lower boundary arc on the unit circle. On this arc it is well-known that the modular invariant $j(\tau)=E_4(\tau)^3/\Delta(\tau)$ is real and ranges from $0$ to $1728$ \cite[Chapter 3]{A}. Let $\alpha$ be a finite zero of $P_m$. Since $P_m(0)=1$, one has $\alpha\ne0$. Choose $\tau$ in the standard fundamental domain with $j(\tau)=\frac{1}{\alpha}$. Then $E_4(\tau)\ne0$ and
\[
 \widehat\Theta_L(\tau)=E_4(\tau)^mP_m(\alpha)=0.
\]
Hence $\tau$ lies on the lower boundary arc, so $0<j(\tau)\le1728$. It follows that every zero $\alpha$ of $P_m$ satisfies $\alpha\ge\frac{1}{1728}$. Counting the zeros with multiplicity and using $P_m(0)=1$, we may therefore write
\[
 P_m(X)=\prod_i\left(1-\frac{X}{\alpha_i}\right),\ \alpha_i\ge\frac{1}{1728}.
\]
Consequently, for $0\le X\le1/1728$,
\[
0\le P_m(X)\le1.
\]

Now take $\tau=iy$ with $y>0$ and consider the auxiliary variable
\[
 u=\frac{\vartheta_2(\tau)^4\vartheta_4(\tau)^4}{\vartheta_3(\tau)^8}=\frac{\vartheta_2(\tau)^4}{\vartheta_3(\tau)^4}\times\frac{\vartheta_4(\tau)^4}{\vartheta_3(\tau)^4}.
\]
Jacobi's identity in the form
\[
 1=\frac{\vartheta_2(\tau)^4}{\vartheta_3(\tau)^4}+\frac{\vartheta_4(\tau)^4}{\vartheta_3(\tau)^4},
\]
 combined with the AGM inequality gives
\[
 0<u\le\frac14.
\]
To get rid of $E_4$ we use the standard theta identities yielding the two equations
\[
 \frac{E_4(\tau)}{\vartheta_3(\tau)^8}=1-u,\ \frac{\Delta(\tau)}{\vartheta_3(\tau)^{24}}=\frac{u^2}{256}.
\]
Therefore, taking quotients we get
\[
 \frac{\Delta(\tau)}{E_4(\tau)^3}=\frac{u^2}{256(1-u)^3}=:X(u).
\]
The function $X(u)$ is increasing for $0<u\le1/4$, and $X\!\left(\frac14\right)=\frac{1}{1728}$. Hence
\[
 0<X(u)\le\frac{1}{1728}.
\]
Dividing the expression for $\widehat\Theta_L$ by $\vartheta_3(\tau)^{8m}$ gives
\[
 \frac{\widehat\Theta_L(\tau)}{\vartheta_3(\tau)^{8m}}=(1-u)^mP_m(X(u)).
\]
Since $0\le P_m(X(u))\le1$ and $0<u\le1/4$, we obtain
\[
 \frac{\widehat\Theta_L(\tau)}{\vartheta_3(\tau)^{8m}}\le(1-u)^m<1.
\]
Finally, for every $0<q<1$ choose $y>0$ such that $q=e^{-\pi y}$. Then $\widehat\Theta_L(iy)=\Theta_L(q)$ and $\vartheta_3(iy)=\vartheta_3(q)$, so
\[
 \Theta_L(q)<\vartheta_3(q)^n,\ 0<q<1.
\]
This proves the result.
\end{proof}

\subsection{Genus averages of Type II lattices}
\label{subsec:typeII-genus-average}

In the preceding subsection, we established a pointwise
Gaussian-mass inequality for extremal even unimodular
lattices. We now consider a different averaging procedure,
namely averaging theta series over the genus of all
positive-definite even unimodular lattices of a fixed rank.
Unlike the averaging methods for Construction $A_4$ and
Construction $A_6$, this procedure does not depend on a
particular code construction.

The connection between genus averages and Eisenstein series
is classical and follows from Siegel's averaging formula.
{It has also been used in the study of secrecy gains of
unimodular lattices (see \cite{sec}). }Here we use this
connection to obtain a uniform pointwise Gaussian-mass
bound for the genus average, with an explicit comparison
to the orthogonal sum of copies of the $E_8$ lattice.

Let $n=8k$, where $k\geq1$, and let $\mathcal{G}_n$
denote the genus of positive-definite even unimodular
lattices of rank $n$. The genus consists of finitely
many isometry classes. Its mass is defined by
\[
\operatorname{Mass}(\mathcal{G}_n)
=
\sum_{[L]\in\mathcal{G}_n}
\frac{1}{|\operatorname{Aut}(L)|}.
\]
The corresponding weighted genus-average theta series is
\[
\Theta_{\mathcal{G}_n}(q)
=
\frac{1}{\operatorname{Mass}(\mathcal{G}_n)}
\sum_{[L]\in\mathcal{G}_n}
\frac{\Theta_L(q)}{|\operatorname{Aut}(L)|}.
\]

To describe this average, let $\tau$ belong to the
complex upper half-plane and put $Q=e^{2\pi i\tau}$. For an integer $r\geq2$, we use the standard normalized
Eisenstein series $E_{2r}(\tau)$ of weight $2r$
(see \cite{A}), defined by
\[
E_{2r}(\tau)
=
1-\frac{4r}{B_{2r}}
\sum_{\ell=1}^{\infty}
\sigma_{2r-1}(\ell)Q^\ell,
\]
where $B_{2r}$ is the Bernoulli number and $\sigma_s(\ell)=\sum_{d\mid\ell}d^s$.

By the classical Siegel genus-average formula
(see also \cite{sec}), the weighted average of
the theta series of even unimodular lattices
satisfies
\[
\widehat{\Theta}_{\mathcal{G}_{8k}}(\tau)
=
E_{4k}(\tau),
\]
where
\[
\widehat{\Theta}_{\mathcal{G}_{8k}}(\tau)
=
\frac{1}{\operatorname{Mass}(\mathcal{G}_{8k})}
\sum_{[L]\in\mathcal{G}_{8k}}
\frac{\widehat{\Theta}_L(\tau)}
{|\operatorname{Aut}(L)|}
\]
and
\[
\widehat{\Theta}_L(\tau)
=
\sum_{x\in L}e^{\pi i\tau(x,x)}.
\]
In particular, if $q=e^{-\pi t}$ with $t>0$, then
\[
\Theta_{\mathcal{G}_{8k}}(q)
=
E_{4k}(it).
\]

We first establish an inequality for normalized
Eisenstein series on the positive imaginary axis.

\begin{lem}
\label{lem:eisenstein-comparison}
For every integer $m\geq2$ and every $t>0$,
\[
|E_{2m}(it)|
\leq
E_4(it)^{m/2}.
\]
Equality holds for $m=2,4$, whereas the inequality
is strict for all other integers $m\geq2$.
In particular, for every integer $k\geq1$,
\[
E_{4k}(it)\leq E_4(it)^k,
\]
with equality if and only if $k=1,2$.
\end{lem}

\begin{proof}
{For $m\geq2$, consider the classical Eisenstein series
(see \cite{A})
\[
G_{2m}(\tau)
=
\sum_{(a,b)\in\Z^2\setminus\{(0,0)\}}
\frac{1}{(a+b\tau)^{2m}}.
\]
}
These series are related to the normalized Eisenstein
series by
\[
G_{2m}(\tau)=2\zeta(2m)E_{2m}(\tau),
\]
where $\zeta$ denotes the Riemann zeta function.

We use the classical recurrence for Eisenstein series,
which follows by comparing coefficients in the Laurent
expansion of the Weierstrass elliptic function (see
\cite{A}). For every integer $m\geq4$,
\[
\begin{aligned}
G_{2m}(\tau)
&=\frac{3}{(m-3)(2m-1)(2m+1)}\\
&\ \ \times\sum_{j=2}^{m-2}(2j-1)(2m-2j-1)
G_{2j}(\tau)G_{2m-2j}(\tau).
\end{aligned}
\]

Substituting
$G_{2r}=2\zeta(2r)E_{2r}$ gives
\[
E_{2m}(\tau)
=
\sum_{j=2}^{m-2}
c_{m,j}E_{2j}(\tau)E_{2m-2j}(\tau),
\]
where
\[
c_{m,j}
=
\frac{
6(2j-1)(2m-2j-1)
\zeta(2j)\zeta(2m-2j)
}{
(m-3)(2m-1)(2m+1)\zeta(2m)
}.
\]
Since $\zeta(s)>0$ for $s>1$, we have $c_{m,j}>0$. Moreover, every normalized Eisenstein series has
constant Fourier coefficient equal to $1$.
Comparing constant coefficients in the preceding
identity yields
\[
\sum_{j=2}^{m-2}c_{m,j}=1.
\]
Thus $E_{2m}$ is a positive convex combination
of products of Eisenstein series of lower weights.

We now prove the desired inequality by strong induction
on $m$.

For $m=2$, the assertion is immediate.

For $m=3$, recall the classical modular-form identity
\[
E_4(\tau)^3-E_6(\tau)^2
=
1728\Delta(\tau),
\]
where
\[
\Delta(\tau)
=
Q\prod_{\ell=1}^{\infty}(1-Q^\ell)^{24}.
\]
When $\tau=it$ with $t>0$, we have
$0<Q<1$, and therefore $\Delta(it)>0$. Since $E_4(it)>0$ and $E_6(it)$ is real,
it follows that $E_6(it)^2<E_4(it)^3$. Consequently,
\[
|E_6(it)|<E_4(it)^{3/2}.
\]

Suppose now that $m\geq4$ and that the assertion
holds for every integer $r$ with $2\leq r<m$.
Using the recurrence and the positivity of
the coefficients $c_{m,j}$, we obtain
\begin{align*}
|E_{2m}(it)|
&\leq
\sum_{j=2}^{m-2}
c_{m,j}
|E_{2j}(it)|
|E_{2m-2j}(it)|\\
&\leq
\sum_{j=2}^{m-2}
c_{m,j}
E_4(it)^{j/2}
E_4(it)^{(m-j)/2}\\
&=
E_4(it)^{m/2}
\sum_{j=2}^{m-2}c_{m,j}\\
&=
E_4(it)^{m/2}.
\end{align*}
This completes the induction.

When $m=4$, the recurrence reduces to $E_8(\tau)=E_4(\tau)^2$, so equality holds. For every $m\geq5$, the recurrence
contains a term involving $E_6$, whose bound is strict
on the positive imaginary axis. Since all the
coefficients $c_{m,j}$ are positive, the resulting
inequality is strict.

Finally, taking $m=2k$ gives
\[
E_{4k}(it)\leq E_4(it)^k,
\]
with equality precisely when $k=1,2$.
\end{proof}

We can now obtain the Gaussian-mass inequality for
the genus average of Type II lattices. We were informed by Prof. Bollauf that a similar result was obtained in \cite{BL}.

\begin{thm}
\label{thm:typeII-genus-average}
Let $n=8k$, where $k\geq1$, and let $\mathcal{G}_n$
denote the genus of positive-definite even unimodular
lattices of rank $n$. Then, for every $0<q<1$,
\[
\Theta_{\mathcal{G}_{8k}}(q)
\leq
\Theta_{E_8^{\oplus k}}(q)
<
\Theta_{\Z^{8k}}(q)
=
\vartheta_3(q)^{8k}.
\]
Equality in the first inequality holds for $k=1,2$,
whereas it is strict for every $k\geq3$.

Moreover, define
\[
u(q)
=
\frac{\vartheta_2(q)^4\vartheta_4(q)^4}
{\vartheta_3(q)^8}.
\]
Then $0<u(q)\leq\frac14$ and
\[
\frac{\Theta_{\mathcal{G}_{8k}}(q)}
{\vartheta_3(q)^{8k}}
\leq
(1-u(q))^k<1.
\]
\end{thm}

\begin{proof}
Fix $0<q<1$ and choose $t>0$ such that $q=e^{-\pi t}$. By the Siegel genus-average formula,
\[
\Theta_{\mathcal{G}_{8k}}(q)
=
E_{4k}(it).
\]
Applying Lemma~\ref{lem:eisenstein-comparison},
we obtain
\[
\Theta_{\mathcal{G}_{8k}}(q)
\leq E_4(it)^k.
\]

The theta series of the $E_8$ lattice is
the normalized Eisenstein series $E_4$ ({see \cite[Chap.~4, Sec.~8.1, p.~123]{SPLAG}).}
Since theta series are multiplicative under
orthogonal direct sums,
\[
\Theta_{E_8^{\oplus k}}(q)
=
\Theta_{E_8}(q)^k
=
E_4(it)^k.
\]
Therefore,
\[
\Theta_{\mathcal{G}_{8k}}(q)
\leq
\Theta_{E_8^{\oplus k}}(q).
\]

To compare this expression with the theta series
of the integer lattice, recall Jacobi's identity
\[
\vartheta_3(q)^4
=
\vartheta_2(q)^4+\vartheta_4(q)^4.
\]
The standard theta-function representation of $E_4$
(equivalently, the theta series of $E_8$; {see \cite[Chap.~4, Sec.~8.1, p.~123]{SPLAG})}
gives
\[
E_4(it)
=
\frac12
\left(
\vartheta_2(q)^8+
\vartheta_3(q)^8+
\vartheta_4(q)^8
\right).
\]
Using Jacobi's identity, this simplifies to
\[
E_4(it)
=
\vartheta_3(q)^8
-
\vartheta_2(q)^4\vartheta_4(q)^4.
\]
Hence
\[
\frac{E_4(it)}{\vartheta_3(q)^8}
=
1-u(q).
\]
{
For $0<q<1$, the defining series show that
$\vartheta_2(q)>0$ and $\vartheta_3(q)>0$.
For $\vartheta_4$, the Jacobi triple-product formula
(see \cite[Chap.~4, Sec.~4.1, pp.~102--103]{SPLAG})) gives
\[
\vartheta_4(q)
=
\prod_{j=1}^{\infty}
(1-q^{2j})(1-q^{2j-1})^2>0.
\]
}
Indeed, all factors are positive and the infinite product
converges to a nonzero limit for $0<q<1$.
Consequently, $u(q)>0$. Furthermore,
\begin{align*}
u(q)
&=
\frac{\vartheta_2(q)^4}{\vartheta_3(q)^4}
\frac{\vartheta_4(q)^4}{\vartheta_3(q)^4}
\leq
\frac14
\left(
\frac{\vartheta_2(q)^4+\vartheta_4(q)^4}
{\vartheta_3(q)^4}
\right)^2=\frac14.
\end{align*}

Combining the preceding identities yields
\begin{align*}
\frac{\Theta_{\mathcal{G}_{8k}}(q)}
{\vartheta_3(q)^{8k}}
&\leq
\frac{E_4(it)^k}{\vartheta_3(q)^{8k}}=(1-u(q))^k<1.
\end{align*}
Therefore,
\[
\Theta_{\mathcal{G}_{8k}}(q)
\leq
\Theta_{E_8^{\oplus k}}(q)
<
\vartheta_3(q)^{8k}.
\]

The equality cases follow from
Lemma~\ref{lem:eisenstein-comparison}.
\end{proof}

The preceding theorem also provides a quantitative
lower bound for the difference between the Gaussian
masses of the integer lattice and the genus average.
Indeed, for every $0<q<1$,
\[
\begin{aligned}
\vartheta_3(q)^{8k}-\Theta_{\mathcal{G}_{8k}}(q)\geq\vartheta_3(q)^{8k}\left(1-(1-u(q))^k\right)>0.
\end{aligned}
\]

At the symmetry point $q=e^{-\pi}$, the classical
identities for Jacobi theta constants imply
\[
\vartheta_2(e^{-\pi})^4
=
\vartheta_4(e^{-\pi})^4
=
\frac12\vartheta_3(e^{-\pi})^4.
\]
Consequently, $u(e^{-\pi})=\frac14$, and Theorem~\ref{thm:typeII-genus-average} gives
\[
\frac{\Theta_{\mathcal{G}_{8k}}(e^{-\pi})}
{\vartheta_3(e^{-\pi})^{8k}}
\leq
\left(\frac34\right)^k.
\]
Equivalently,
\[
\frac{\vartheta_3(e^{-\pi})^{8k}}
{\Theta_{\mathcal{G}_{8k}}(e^{-\pi})}
\geq
\left(\frac43\right)^k.
\]

We conclude by emphasizing the distinction between
this result and an individual Gaussian-mass inequality.
Theorem~\ref{thm:typeII-genus-average} concerns the
weighted average of the theta series over the full
genus. It does not, by itself, imply that every
even unimodular lattice $L$ of rank $8k$ satisfies $\Theta_L(q)\leq\vartheta_3(q)^{8k}$ for all $0<q<1$.

Nevertheless, the genus-average bound holds in every
rank divisible by eight, without requiring a
classification of the lattices in the genus or
a restriction on their minimum norms. It therefore
provides a complementary averaged perspective to
the pointwise results for extremal Type II lattices.

\section{Extremal Type I lattices}
\label{sec:typeI}

We now prove the strict Gaussian-mass inequality for extremal odd
unimodular lattices. The argument uses the shadow and applies without
an $s$-extremality assumption.

\subsection{Statement and scope}
For an odd unimodular lattice $L$, let
\[
L_0=\{x\in L:(x,x)\in2\mathbb Z\},\
S=L_0^*\setminus L
\]
be its even sublattice and its shadow, respectively
(see \cite[Section~10]{GSZ}), and write $\Theta_S(q)=\sum_{x\in S}q^{(x,x)}$ for the theta series of $S$.
We use the theta constants defined earlier and put
\[
\Delta_8(q)=\frac{\vartheta_2(q)^4\vartheta_4(q)^4}{16}.
\]
In this section, \emph{extremal} means attaining the Rains--Sloane
minimum-norm bound \cite{RS}: the bound is
$2\lfloor n/24\rfloor+2$ when $n\ne23$, with the exceptional bound
$3$ when $n=23$.

\begin{thm}\label{thm:typeI-main}
Let $L$ be a positive-definite integral odd unimodular lattice of rank $n>0$.
Suppose either $n\ne23,\ \min(L)=2\lfloor n/24\rfloor+2$, or $n=23$ and $\min(L)=3$. Then
\[
 \Theta_L(q)<\vartheta_3(q)^n\ (0<q<1).
\]
Equivalently,
\[
\sum_{x\in L}e^{-s(x,x)}<\sum_{z\in\mathbb Z^n}e^{-s(z,z)}
\]
for every $s>0$.
\end{thm}

This statement covers extremal Type I lattices in every rank in which they exist.

\subsection{The universal theta truncation}
Set
\[
 u_{\mathrm I}=\frac{\Delta_8}{\vartheta_3^8},\ t=64u_{\mathrm I},
\ K=\lfloor n/8\rfloor.
\]
The standard theta and shadow expansions \cite{RS} are
\begin{align}
 \Theta_L(q)&=\sum_{j=0}^{K}c_j\Delta_8(q)^j\vartheta_3(q)^{n-8j},\label{eq:typeI-lattice}\\
 \Theta_S(q)&=\sum_{j=0}^{K}\frac{(-1)^jc_j}{16^j}
 \vartheta_4(q^2)^{8j}\vartheta_2(q)^{n-8j}.\label{eq:typeI-shadow}
\end{align}
{
The theta series lies in the graded algebra generated by
$\vartheta_3$ and $\Delta_8$. Its weight-$n/2$ component has basis
$\Delta_8^j\vartheta_3^{n-8j}$, for $0\le j\le K$.
Here $c_j=c_j(L)$, for $0\le j\le K$, denotes the uniquely determined
coefficient in the theta-series expansion on this basis \eqref{eq:typeI-lattice}.
In particular, $c_0=1$. The same coefficients occur in the shadow
expansion \eqref{eq:typeI-shadow}.
}
Thus
\[
 P(t):=\frac{\Theta_L(q)}{\vartheta_3(q)^n}
      =\sum_{j=0}^{K}\frac{c_j}{64^j}t^j.
\]
The parameter $u_{\mathrm I}$ is $1/16$ times the parameter $u$
used in Section~\ref{sec:typeII}. Thus the theta-function identities
recalled in Subsection~\ref{subsec:typeII-genus-average} give
\[
0<t(q)=4\frac{\vartheta_2(q)^4\vartheta_4(q)^4}
                  {\vartheta_3(q)^8}\le1,
\ t(q_0)=1,\ q_0=e^{-\pi}.
\]
{
To specify the hypergeometric identity and its branch, put
\[
 \lambda(q)=\frac{\vartheta_2(q)^4}{\vartheta_3(q)^4},\
 1-\lambda(q)=\frac{\vartheta_4(q)^4}{\vartheta_3(q)^4}.
\]
Then $t(q)=4\lambda(q)(1-\lambda(q))$. The classical quadratic transformation
\cite[Chap.~15]{NIST}, with $a=b=1/2$, reads
\[
\begin{gathered}
{}_2F_1(\tfrac12,\tfrac12;1;\lambda)
={}_2F_1(\tfrac14,\tfrac14;1;4\lambda(1-\lambda)),\  \operatorname{Re}\lambda<\tfrac12.
\end{gathered}
\]
Also, the elliptic-integral identities
\cite[Chaps.~19--20]{NIST} give
\[
 {}_2F_1(\tfrac12,\tfrac12;1;\lambda(q))
 =\frac{2}{\pi}\mathrm K(\sqrt{\lambda(q)})=\vartheta_3(q)^2,
\]
where $\mathrm K$ denotes the complete elliptic integral of the first kind.
Consequently, on the branch at $q=0$,
\[
 F(t):={}_2F_1(\tfrac14,\tfrac14;1;t),\
 F(t(q))=\vartheta_3(q)^2.
\]
Only this local, equivalently formal-power-series, identity is used to
determine coefficients. It is not asserted on the same branch for all
$0<q<1$.
}
Write
\[
 b_j=\coeff{t}{j}F(t)^{-n/2},\ Q_{n,d}(t)=\sum_{j=0}^d b_jt^j.
\]
{
Suppose $\min(L)=\mu$ and $d=\mu-1\le K$. The minimum-norm
condition says precisely that
\[
 \Theta_L(q)=1+O(q^\mu).
\]
Since $F(t(q))^{-n/2}=\vartheta_3(q)^{-n}$, division by the unit series
$\vartheta_3(q)^n$ gives
\[
 P(t(q))-F(t(q))^{-n/2}
 =\frac{\Theta_L(q)-1}{\vartheta_3(q)^n}=O(q^\mu).
\]
Now $\Delta_8(q)=q+O(q^2)$ and $\vartheta_3(q)=1+O(q)$, so
$t(q)=64q+O(q^2)$. Hence there is a formal inverse
$q(t)=t/64+O(t^2)$, and substitution yields
\[
 P(t)-F(t)^{-n/2}=O(t^\mu).
\]
Comparing coefficients of $t^j$, using the definitions of $P$ and $b_j$,
we obtain
\[
 \frac{c_j}{64^j}=b_j\ (0\le j<\mu).
\]
This proves the decomposition
\begin{equation}\label{eq:typeI-decomp}
 P(t)=Q_{n,d}(t)+\sum_{j=\mu}^{K}\frac{c_j}{64^j}t^j.
\end{equation}

For clarity, the vanishing coefficients are $\coeff{q}{j}\Theta_L(q)$
for $1\le j<\mu$, not the basis coefficients $c_j$.
Although $\Delta_8^j\vartheta_3^{n-8j}=q^j+O(q^{j+1})$, the
$j=0$ term is $\vartheta_3^n$, not the constant series $1$.
For example,
\[
\begin{aligned}
\vartheta_3(q)^n=1+2nq+2n(n-1)q^2+O(q^3), \ \Delta_8(q)=q-8q^2+O(q^3).
\end{aligned}
\]
Thus $\mu\ge2$ forces $2n+c_1=0$, or $c_1=-2n$.
If $\mu\ge3$, the coefficient of $q^2$ similarly gives
\[
 2n(n-1)+(2n-24)c_1+c_2=0,
\ c_2=2n(n-23).
\]
These agree with $b_1=-n/32$ and $b_2=n(n-23)/2048$ through
$c_j=64^j b_j$.
}

\begin{lem}\label{lem:typeI-bernstein}
Let $n,d$ be positive integers with $n\le12d+15$. There are strictly positive
numbers $g_0,\ldots,g_{d-1}$ such that
\begin{equation}\label{eq:typeI-bernstein}
 -Q_{n,d}'(t)=\sum_{i=0}^{d-1}g_i t^i(1-t)^{d-1-i}.
\end{equation}
Consequently $Q_{n,d}'(t)<0$ on $[0,1]$, and
\begin{equation}\label{eq:typeI-deficit}
 1-Q_{n,d}(t)\ge t^d\bigl(1-Q_{n,d}(1)\bigr),\ 0\le t\le1.
\end{equation}
\end{lem}

\begin{proof}
Define
\[
 g_i=\coeff{w}{i}(1+w)^{d-1}
       [-Q_{n,d}'(w/(1+w))].
\]
{
Here $0\le i\le d-1$. Set $H(t)=F(t)^{-n/2}$.
Since $H(t)-Q_{n,d}(t)=O(t^{d+1})$, we have
$H'(t)-Q_{n,d}'(t)=O(t^d)$. Substitution of $t=w/(1+w)$
and multiplication by $(1+w)^{d-1}$ preserve order at least $d$.
Thus $Q_{n,d}'$ may be replaced by $H'$ when computing $g_i$.
Writing coefficient extraction as a formal residue gives
\[
 g_i=-\operatorname*{Res}_{w=0}
 (1+w)^{d-1}H'\!\left(\frac{w}{1+w}\right)
 w^{-i-1}\,dw.
\]
Under $t=w/(1+w)$, we have
\[
 w=\frac{t}{1-t},\ 1+w=\frac{1}{1-t},\
 dw=\frac{dt}{(1-t)^2}.
\]
In particular,
\[
 (1+w)^{d-1}w^{-i-1}\,dw
 =t^{-i-1}(1-t)^{i-d}\,dt,
\]
and therefore
\[
 g_i=-\operatorname*{Res}_{t=0}
 H'(t)t^{-i-1}(1-t)^{i-d}\,dt.
\]
Next substitute $t=t(q)$. Both changes of variable are valid for
formal residues, because their linear coefficients at zero are nonzero.
The chain rule gives the differential identity
\[
\begin{aligned}
H'(t(q))\,dt&=H'(t(q))t'(q)\,dq
=\frac{d}{dq}\bigl(\vartheta_3(q)^{-n}\bigr)\,dq
=-n\vartheta_3'(q)\vartheta_3(q)^{-n-1}\,dq.
\end{aligned}
\]
Equivalently, the factor $t'(q)$ from $dt=t'(q)\,dq$ cancels
the factor $1/t'(q)$ in the chain-rule expression for $H'(t(q))$.
No derivative factor is omitted. We obtain
\[
 g_i=n\operatorname*{Res}_{q=0}
 \vartheta_3'(q)\vartheta_3(q)^{-n-1}
 t(q)^{-i-1}(1-t(q))^{i-d}\,dq.
\]
Finally, the Jacobi product identities give
\[
\begin{gathered}
\Phi(q)=\prod_{m\ge1}(1+q^{2m-1}),
u_{\mathrm I}(q)=q\Phi(q)^{-24},\ t(q)=64q\Phi(q)^{-24}.
\end{gathered}
\]
Using $\operatorname*{Res}_{q=0}q^{-i-1}A(q)\,dq
=\coeff{q}{i}A(q)$ yields
\begin{equation}
\label{eq:typeI-g-residue}
\begin{aligned}
g_i={}\frac{n}{64^{i+1}}\coeff{q}{i}
\vartheta_3'(q)\vartheta_3(q)^{-n-1}
\times\Phi(q)^{24(i+1)}(1-t(q))^{-(d-i)},
\end{aligned}
\end{equation}
where $\vartheta_3'(q)$ denotes differentiation with respect to $q$;
$H'$ and $Q_{n,d}'$ denote differentiation with respect to $t$.
}

Put $k=d-i\ge1$, $R=\vartheta_2^4/\vartheta_4^4$, and
$E=16d+8i+22-2n$. Since
\[
 1-t=\frac{\vartheta_4^8}{\vartheta_3^8}(1-R)^2,
\]
the series inside \eqref{eq:typeI-g-residue} is
\begin{equation}
\label{eq:typeI-g-product}
\begin{aligned}
\vartheta_3'(q)(1-R(q))^{-2k}
\prod_{m\ge1}(1-q^{2m})^{-n-1}
\times\prod_{m\ge1}(1+q^{2m-1})^E(1-q^{2m-1})^{-16k}.
\end{aligned}
\end{equation}
Both $\vartheta_3'$ and $R$ have nonnegative coefficients, and $R=16q+O(q^2)$.
If $E\ge0$, every factor in \eqref{eq:typeI-g-product} has nonnegative coefficients.
If $E<0$, use
\[
 (1+z)^E(1-z)^{-16k}=(1-z^2)^E(1-z)^{-(16k+E)}.
\]
Here
\[
 16k+E=32d-8i+22-2n\ge24d+30-2n\ge0.
\]
Again all coefficients are nonnegative. The constant term $\vartheta_3'(0)=2$
and the positive coefficient of every power in $(1-R)^{-2k}$ make $g_i>0$.
This proves \eqref{eq:typeI-bernstein}. Since $g_i>0$, the right-hand side
of \eqref{eq:typeI-bernstein} is positive for $0<t<1$; at $t=0$ and $t=1$
it equals $g_0$ and $g_{d-1}$, respectively. Hence $Q_{n,d}'<0$
on the whole interval $[0,1]$.

We now prove \eqref{eq:typeI-deficit} in detail. Write $Q=Q_{n,d}$.
Since $F(0)=1$, we have $Q(0)=b_0=1$. Integrating
\eqref{eq:typeI-bernstein} from $0$ to $t$ gives
\[
\begin{gathered}
1-Q(t)=\sum_{i=0}^{d-1}g_i I_i(t),\\
I_i(t):=\int_0^t x^i(1-x)^{d-1-i}\,dx.
\end{gathered}
\]
Fix $0<t<1$ and $0\le i\le d-1$. The substitution $x=ts$ yields
\[
 I_i(t)=t^{i+1}\int_0^1 s^i(1-ts)^{d-1-i}\,ds.
\]
For $0\le s\le1$, we have $1-ts\ge1-s\ge0$.
Because $d-1-i$ is a nonnegative integer, it follows that
$(1-ts)^{d-1-i}\ge(1-s)^{d-1-i}$. Moreover,
$i+1\le d$ and $0<t<1$ imply $t^{i+1}\ge t^d$.
Consequently,
\begin{align*}
 I_i(t)
 \ge t^{i+1}\int_0^1s^i(1-s)^{d-1-i}\,ds
 \ge t^d\int_0^1s^i(1-s)^{d-1-i}\,ds
 =t^d I_i(1).
\end{align*}
Multiplying these inequalities by $g_i>0$ and summing over $i$,
we obtain
\[
\begin{aligned}
1-Q(t)=\sum_{i=0}^{d-1}g_iI_i(t)
\ge t^d\sum_{i=0}^{d-1}g_iI_i(1)
=t^d\bigl(Q(0)-Q(1)\bigr)
=t^d\bigl(1-Q(1)\bigr).
\end{aligned}
\]
At $t=0$ both sides are zero, and at $t=1$ equality holds.
This proves \eqref{eq:typeI-deficit} for every $0\le t\le1$.
\end{proof}

\subsection{Exact inversion of the shadow coefficients}
Write $n=8K+r_8$, $0\le r_8\le7$, and set $x=q^2$.
The shadow has an expansion
\[
 \Theta_S(q)=q^{r_8/4} B(x),\
 B(x)=\sum_{a\ge0}B_ax^a,\ B_a\ge0.
\]
Define the following convergent products for $|x|<1$:
\begin{align*}
T(x)&=\sum_{a\ge0}x^{a(a+1)/2}
=\prod_{m\ge1}\frac{1-x^{2m}}{1-x^{2m-1}},\\
C(x)&=\vartheta_4(x)=\prod_{m\ge1}(1-x^{2m})(1-x^{2m-1})^2,\\
w(x)&=x(T(x)/C(x))^8
=x\prod_{m\ge1}(1-x^{2m-1})^{-24},\\
A(x)&=\frac{xw'(x)}{w(x)}
=1+24\sum_{\substack{a\ge1\\a\text{ odd}}}\frac{a x^a}{1-x^a},\\
H_j(x)&=C(x)^{-8j}T(x)^{-(n-8j)}\\
&=\prod_{m\ge1}(1-x^{2m})^{-n}(1-x^{2m-1})^{n-24j}.
\end{align*}

\begin{lem}\label{lem:typeI-shadow}
For $0\le j\le K$,
\begin{equation}\label{eq:typeI-inverse}
 (-1)^j2^{n-12j}c_j=\coeff{x}{K-j}B(x)H_j(x)A(x).
\end{equation}
In particular, if $24j\ge n$, then $(-1)^jc_j\ge0$ and
\begin{equation}
\label{eq:typeI-c-bound}
\begin{gathered}
\frac{|c_j|}{64^j}\le A_0\,2^{n/4}8^{-j}P(1),
A_0=A(e^{-2\pi})<\frac{21}{20}.
\end{gathered}
\end{equation}
\end{lem}

\begin{proof}
Using $\vartheta_2(q)=2q^{1/4}T(q^2)$, equation \eqref{eq:typeI-shadow} becomes
\[
\begin{aligned}
B(x)={}2^{n-12K}C(x)^{8K}T(x)^{r_8}
\times\sum_{k=0}^{K}(-1)^{K-k}c_{K-k}\,4096^k w(x)^k.
\end{aligned}
\]
The formal coefficient change of variable, with $w=x f(x)$, is
\[
 \coeff{w}{k}G(x(w))
 =\coeff{x}{k}G(x)f(x)^{-k}\frac{xw'(x)}{w(x)}.
\]
Apply this with $k=K-j$ and $f=(T/C)^8$ to obtain \eqref{eq:typeI-inverse}.
If $24j\ge n$, $B,H_j,A$ all have nonnegative coefficients.
{
We use the following elementary coefficient estimate. If
$G(x)=\sum_{\ell\ge0}g_\ell x^\ell$ has nonnegative coefficients
and converges at $x_0>0$, then
\[
 [x^m]G(x)=g_m\le x_0^{-m}G(x_0),
\]
because $G(x_0)=\sum_{\ell\ge0}g_\ell x_0^\ell\ge g_m x_0^m$.
Apply this estimate to $G(x)=B(x)H_j(x)A(x)$ with
$m=K-j$ and $x_0=q_0^2=e^{-2\pi}$. Using \eqref{eq:typeI-inverse}, we get
\[
 |c_j|\le 2^{12j-n}x_0^{-(K-j)}B(x_0)H_j(x_0)A_0,
\ A_0=A(x_0).
\]
}
After replacing $B,T,C$ by theta functions, this is
\[
 \frac{|c_j|}{64^j}
 \le A_0\,\Theta_S(q_0)\vartheta_2(q_0)^{-n}
 \left(\frac{\vartheta_2(q_0)^8}{4\vartheta_4(q_0^2)^8}\right)^j.
\]
At $q_0=e^{-\pi}$, the Jacobi transformation and duplication identities give
\[
\begin{gathered}
\vartheta_2(q_0)^4=\vartheta_4(q_0)^4=\tfrac12\vartheta_3(q_0)^4,
\vartheta_4(q_0^2)^2=\vartheta_3(q_0)\vartheta_4(q_0).
\end{gathered}
\]
Thus the parenthesis is $1/8$ and
$\vartheta_2(q_0)^{-n}=2^{n/4}\vartheta_3(q_0)^{-n}$.

{
The shadow is one translate $v+L$ of $L$, since $[L_0^*:L]=2$ and
$S=L_0^*\setminus L$. We recall the Gaussian translate inequality
of Micciancio and Regev \cite[Lemma~2.9]{MR}, including its derivation
from their Poisson summation formula \cite[Lemma~2.8]{MR}.
Use the Fourier-transform convention
\[
 \widehat f(z)=\int_{\mathbb R^n}f(x)e^{-2\pi i(x,z)}\,dx.
\]
For a Schwartz function $f$ and $g(x)=f(x+v)$, the substitution
$r=x+v$ gives $\widehat g(z)=e^{2\pi i(z,v)}\widehat f(z)$.
Applying Poisson summation to $g$ on a full-rank lattice $M$ yields
\[
 \sum_{x\in M+v}f(x)
 =\frac{1}{\operatorname{covol}(M)}
   \sum_{z\in M^*}e^{2\pi i(z,v)}\widehat f(z).
\]
For $f_y(x)=e^{-\pi y\|x\|^2}$, $y>0$, its Fourier transform is
$\widehat f_y(z)=y^{-n/2}e^{-\pi\|z\|^2/y}$.
Pairing the terms indexed by $z$ and $-z$, and then using
$\cos(2\pi(z,v))\le1$, gives
\[
\begin{aligned}
\sum_{x\in M+v}e^{-\pi y\|x\|^2}
&=\frac{y^{-n/2}}{\operatorname{covol}(M)}
\sum_{z\in M^*}e^{-\pi\|z\|^2/y}\cos(2\pi(z,v))\\
&\le\frac{y^{-n/2}}{\operatorname{covol}(M)}
\sum_{z\in M^*}e^{-\pi\|z\|^2/y}\\
&=\sum_{x\in M}e^{-\pi y\|x\|^2}.
\end{aligned}
\]
The final equality is Poisson summation with $v=0$.
All Gaussian sums here converge absolutely.
}
In particular $\Theta_S(q_0)\le\Theta_L(q_0)$, proving the main estimate.
Finally $x_0<1/500$ and
\[
 A_0\le1+\frac{24x_0}{(1-x_0)^3}
 <1+\frac{24/500}{(499/500)^3}<\frac{21}{20}.
\]
The elementary bound $x_0<1/500$ follows, for example, from
$\pi>25/8$ and $\sum_{a=0}^{16}(25/4)^a/a!>500$.
\end{proof}

\subsection{A uniform bound for the additional terms}
From now on let
\[
 n=24m+r,\ m\ge1,\ 0\le r\le23,\
 \mu=2m+2,\ d=2m+1.
\]
Then $n=12d+c$, where $c=r-12\in\{-12,\ldots,11\}$, and
$n\le12d+15$. Moreover $24j\ge n$ for every $j\ge\mu$.
Since $\mu$ is even, Lemma~\ref{lem:typeI-shadow} implies that the positive terms
in the tail of \eqref{eq:typeI-decomp} have even indices. Discard the nonpositive
odd-index terms and use $t^j\le t^\mu$ for $0\le t\le1$ to obtain
\begin{equation}
\label{eq:typeI-contraction}
\begin{gathered}
P(t)\le Q_{n,d}(t)+\kappa_r t^\mu P(1),
\kappa_r=\frac{A_0}{1-1/64}\,2^{r/4-6}.
\end{gathered}
\end{equation}
This also holds when the tail is empty. In particular,
\begin{equation}\label{eq:typeI-kappa}
 0<\kappa_r<\frac{16}{15}2^{r/4-6}<\frac9{10}.
\end{equation}
For the last inequality, use $r\le23$ and
$2^{1/4}>32/27$, verified by fourth powers.

\begin{prop}\label{prop:typeI-head}
With the above parameters,
\[
Q_{n,d}(1)+\kappa_r<1.
\]
\end{prop}

\begin{proof}
Write $Q=Q_{n,d}$. By Lemma~\ref{lem:typeI-bernstein},
\[
 1-Q(1)=\sum_{i=0}^{d-1}\frac{g_i}{d\binom{d-1}{i}}.
\]
All summands are positive. We use only the first one, two, or five.

If $0\le r\le12$, then $\kappa_r<2/15$, while
\[
 1-Q(1)\ge\frac{g_0}{d}=\frac{n}{32d}
 \ge\frac{12d-12}{32d}\ge\frac14.
\]
If $13\le r\le19$, so $1\le c\le7$, then $\kappa_r<1/2$, and
\[
 \frac{g_0}{d}+\frac{g_1}{d(d-1)}-\frac12
 =\frac{112d^2+40dc+20d-c^2-41c}{1024d(d-1)}>0.
\]
The numerator is positive because $d\ge3$ and $40d-c-41>0$.

It remains to consider $20\le r\le23$, or $8\le c\le11$.
For $d=3$, direct substitution gives
\[
\begin{array}{c|rrrr}
 c&8&9&10&11\\\hline
 Q_{12d+c,d}(1)&213/4096&49/1024&721/16384&1325/32768
\end{array}
\]
and each value is less than $1/10$.
For odd $d\ge5$, put
\[
 I_4=\sum_{i=0}^{4}\frac{g_i}{d\binom{d-1}{i}}.
\]
The exact polynomial calculation in Subsection~\ref{subsec:typeI-certificate} proves
\[
 I_4>1-(5/8)^5>9/10.
\]
Hence $Q(1)<1/10$ in both cases, whereas $\kappa_r<9/10$.
\end{proof}

\subsection{Exact certificate for the five-term estimate}\label{subsec:typeI-certificate}
The first coefficients of $F(t)^{-n/2}$ are
\[
 b_0=1,\ b_1=-\frac n{32},\
 b_2=\frac{n(n-23)}{2048},
\]
\begin{align*}
 b_3&=-\frac{n(n^2-69n+1208)}{196608},\\
 b_4&=\frac{n(n^3-138n^2+6419n-100677)}{25165824},\\
 b_5&=-\frac{n(n^4-230n^3+20015n^2-781225n+11542704)}{4026531840}.
\end{align*}
For $0\le i\le4$,
\[
 g_i=-\sum_{j=1}^{i+1}j b_j\binom{d-j}{i+1-j}.
\]
Set $x=d-5\ge0$. With $n=12d+c$ and $c=8,9,10,11$, exact simplification gives
\[
 I_4-\bigl(1-(5/8)^5\bigr)
 =\frac{R_c(x)}{33554432\,d(d-1)(d-2)(d-3)(d-4)},
\]
where
\begin{align*}
 R_8(x)={}&16960000x^4+162644800x^3+547478960x^2\\
          &+756889760x+355317360,\\
 R_9(x)={}&17760000x^4+167620800x^3+557975520x^2\\
          &+765530880x+357599985,\\
 R_{10}(x)={}&18560000x^4+172436800x^3+567910480x^2\\
             &+773570720x+359702340,\\
 R_{11}(x)={}&19360000x^4+177092800x^3+577307840x^2\\
             &+781046960x+361639185.
\end{align*}
Every coefficient is positive. This is an exact proof for all $d\ge5$, not
a check up to a finite dimension. The separate $d=3$ values in the proof are
the only small-degree exceptions needed.

\subsection{Completion of the proof}
For $n\ge24$, equation \eqref{eq:typeI-contraction} at $t=1$ gives
\[
 (1-\kappa_r)P(1)\le Q(1)<1-\kappa_r,
\]
so $P(1)<1$. Notice that $P(1)>0$ also forces $Q(1)>0$ whenever the
assumed lattice exists; no independent existence or positivity assertion about
the formal truncation is needed.
For every $0<t\le1$, Lemma~\ref{lem:typeI-bernstein} and
Proposition~\ref{prop:typeI-head} give
\begin{align*}
 P(t)\le Q(t)+\kappa_r t^{d+1}P(1)
     < Q(t)+\kappa_r t^d
     < Q(t)+t^d(1-Q(1))\le1.
\end{align*}
This is the desired dimension-independent estimate.

For completeness the small ranks can be treated without enumerating lattices.
If $n<8$, equation \eqref{eq:typeI-lattice} forces $\Theta_L=\vartheta_3^n$, so no
lattice with minimum $2$ exists. If $8\le n\le15$ and $\min(L)=2$, then

\[
P(t)=1-nt/32<1.
\]
For $16\le n\le22$ and $\min(L)=2$, write $N_2$ for the number of roots.
The ADE root-system formulas imply $N_2\le2n(n-1)$: every irreducible component
has root-to-rank ratio at most $2(n-1)$, using $30\le2(n-1)$ for the exceptional
components. Comparing the coefficient of $q^2$ gives
\[
 c_1=-2n,\ c_2=N_2-2n(23-n)\le4n(n-12).
\]
Consequently
\[
\begin{aligned}
P(t)\le1-\frac n{32}t+\frac{n(n-12)}{1024}t^2
\le1-\frac{n(44-n)}{1024}t<1.
\end{aligned}
\]
For $n=23$ and minimum $3$, the same coefficient comparison with $N_2=0$
gives $c_2=0$, hence
\[
P(t)=1-23t/32<1.
\]
This completes
Theorem~\ref{thm:typeI-main}.

\begin{rem}
The estimate \eqref{eq:typeI-contraction} uses the extremal minimum norm.
Consequently, the argument establishes the stated inequality for extremal
Type~I lattices; it does not establish it for arbitrary nonextremal odd
unimodular lattices.
\end{rem}

\section{Higher genus}
\label{sec:higher-genus}
\subsection{Binary case}
In this section we generalize the Kominers inequality to higher genus theta series.
 Let $C\subseteq\F_2^{2k}$ be a binary self-dual code with generator matrix
\[
G=[I_k\mid A],
\  AA^T=I.
\]

For \(\varepsilon\in\F_2^g\), define
\[
\psi_\varepsilon(Y)
=
\sum_{m\in\Z^g}
\exp\left[
-\frac{\pi}{2}
(\varepsilon+2m)^TY(\varepsilon+2m)
\right].
\]
For \(U\in\F_2^{g\times k}\), let $F(U)=\prod_{j=1}^k\psi_{U_j}(Y)$. Then
\[
G_{A_2(C)}^{(g)}(Y)
=
\sum_U F(U)F(UA).
\]
By the Cauchy--Schwarz inequality,
\[
G_{A_2(C)}^{(g)}(Y)
\le
\sum_U F(U)^2.
\]
On the other hand, the change of variables $p=\frac{u+v}{2}, \ q=\frac{u-v}{2}$ gives
\[
\sum_{\varepsilon\in\F_2^g}
\psi_\varepsilon(Y)^2
=
\Theta_g(Y)^2.
\]
Therefore
\begin{equation}
{
G_{A_2(C)}^{(g)}(Y)
\le
\Theta_g(Y)^{2k}
=
G_{\Z^{2k}}^{(g)}(Y)
}
\label{eq:hg-binary-bound}
\end{equation}
for every genus \(g\).
{
\subsection{Quaternary case}
For a lattice $L\subseteq\mathbb R^n$ and a real symmetric positive-definite
matrix $Y\in\operatorname{Sym}_g^+(\mathbb R)$, write
\begin{equation}
G_L^{(g)}(Y)
=\sum_{x_1,\ldots,x_g\in L}
\exp\!\left(-\pi\sum_{a,b=1}^gY_{ab}\langle x_a,x_b\rangle\right).
\label{eq:hg-gaussian-definition}
\end{equation}
This is the genus-$g$ Siegel theta series of $L$ evaluated at $iY$.
Set
\begin{equation}
\Theta_g(Y)=\sum_{z\in\mathbb Z^g}\exp(-\pi z^TYz).
\label{eq:hg-standard-theta}
\end{equation}
Then $G_{\mathbb Z^n}^{(g)}(Y)=\Theta_g(Y)^n$.
For a self-dual code $C\subseteq\mathbb Z_4^n$, its Construction~A lattice is
\[
A_4(C)=\frac12\{z\in\mathbb Z^n:z\bmod4\in C\}.
\]
We use $R=\operatorname{Res}(C),\ k=\dim_{\mathbb F_2}R$ only as notation; throughout, $R$ is doubly even and
$\operatorname{Tor}(C)=R^\perp$.
For $1\le s\le k$, the $s$th generalized Hamming weight of $R$ is
\begin{equation}
d_s(R)=\min\{\#\operatorname{supp}(U):U\le R,\ \dim U=s\},
\label{eq:hg-generalized-weight}
\end{equation}
where $\operatorname{supp}(U)=\bigcup_{u\in U}\operatorname{supp}(u)$.

We first record the theta identities needed below. Fix $g\ge1$ and $Y>0$,
and put $d=2^g$ and $\Theta=\Theta_g(Y)$. For
$\varepsilon,\eta\in\mathbb F_2^g$, define
\begin{equation}
B_{\varepsilon,\eta}(Y)
=\sum_{z\in\mathbb Z^g}(-1)^{\eta\cdot z}
\exp\!\left[-\pi(z+\varepsilon/2)^TY(z+\varepsilon/2)\right].
\label{eq:hg-characteristic-theta}
\end{equation}
Let $\mathcal H=\mathbb F_2^g\times\mathbb F_2^g$ and write
$b_{\varepsilon,\eta}=B_{\varepsilon,\eta}/\Theta$.

Choose a section $\sigma:R\to C$ of reduction modulo $2$, and write
$\sigma(r)=r+2h(r)$ with $h(r)\in\mathbb F_2^n$.
Every codeword of $C$ is uniquely expressible as $\sigma(r)+2t$,
where $r\in R$ and $t\in R^\perp$.
For $\mathbf r=(r_1,\ldots,r_g)\in R^g$ and
$\mathbf s=(s_1,\ldots,s_g)\in R^g$, set
\[
\varepsilon_i=(r_{1i},\ldots,r_{gi}),\
\eta_i=(s_{1i},\ldots,s_{gi}).
\]
Character orthogonality applied to each torsion row gives
\begin{equation}
G_{A_4(C)}^{(g)}(Y)
=d^{-k}\sum_{\mathbf r,\mathbf s\in R^g}
(-1)^{\sum_{j=1}^g s_j\cdot h(r_j)}
\prod_{i=1}^nB_{\varepsilon_i,\eta_i}(Y).
\label{eq:hg-fourier-expansion}
\end{equation}
Indeed, the projector onto $R^\perp$ is
$2^{-k}\sum_{s\in R}(-1)^{s\cdot t}$.
After translating the torsion coordinate by $h(r)$, its local Fourier
transform is exactly \eqref{eq:hg-characteristic-theta}.

\begin{lem}
\label{lem:hg-quartic}
For every $g\ge1$ and $Y\in\operatorname{Sym}_g^+(\mathbb R)$,
\begin{equation}
|b_x|\le1,\ \sum_{x\in\mathcal H}|b_x|^4=d.
\label{eq:hg-quartic-normalization}
\end{equation}
\end{lem}

\begin{pf}
Removing the signs in \eqref{eq:hg-characteristic-theta} gives
\[
|B_{\varepsilon,\eta}|
\le\sum_{z\in\mathbb Z^g}
\exp\!\left[-\pi(z+\varepsilon/2)^TY(z+\varepsilon/2)\right]
\le\Theta.
\]
The last inequality follows from Poisson summation: the periodized
Gaussian has positive Fourier coefficients and attains its maximum at
an integer shift.

For the quartic identity, let
$E=\{t\in\mathbb F_2^4:t_1+t_2+t_3+t_4=0\}$ and consider the self-dual
code
\[
C_4=\langle(1,1,1,1),2E\rangle_{\mathbb Z_4}.
\]
Its lattice is
\[
A_4(C_4)=D_4\cup\bigl(D_4+\tfrac12(1,1,1,1)\bigr),
\]
where $D_4=\{z\in\mathbb Z^4:\sum_i z_i\equiv0\pmod2\}$.
The rows of $H/2$, with
\begin{equation}
H=\begin{pmatrix}
1&1&1&1\\
1&1&-1&-1\\
1&-1&1&-1\\
1&-1&-1&1
\end{pmatrix},\ HH^T=4I_4,
\label{eq:hg-hadamard}
\end{equation}
belong to this lattice and form an orthonormal basis. Since the lattice
is unimodular, they generate it. Thus $A_4(C_4)\cong\mathbb Z^4$.
Taking the section $\sigma(1,1,1,1)=(1,1,1,1)$ in
\eqref{eq:hg-fourier-expansion} yields
\[
d^{-1}\sum_{\varepsilon,\eta\in\mathbb F_2^g}
B_{\varepsilon,\eta}^4=\Theta^4.
\]
The quantities $B_{\varepsilon,\eta}$ are real, so division by $\Theta^4$
proves the second assertion.
\end{pf}

\begin{thm}
\label{thm:hg-generalized-weights}
Let $C\subseteq\mathbb Z_4^n$ be self-dual, and let
$R=\operatorname{Res}(C)$ have dimension $k$.
If
\begin{equation}
d_s(R)\ge4s\,\text{for every }1\le s\le k,
\label{eq:hg-weight-condition}
\end{equation}
then, for every $g\ge1$ and every $Y\in\operatorname{Sym}_g^+(\mathbb R)$,
\begin{equation}
G_{A_4(C)}^{(g)}(Y)\le\Theta_g(Y)^n.
\label{eq:hg-generalized-weight-bound}
\end{equation}
\end{thm}

\begin{pf}
The assertion is immediate when $k=0$, since then
$C=2\mathbb Z_4^n$ and $A_4(C)=\mathbb Z^n$.
Assume $k>0$ and choose a generator matrix
$G_R=(v_1,\ldots,v_n)$ with columns $v_i\in\mathbb F_2^k$.

We claim that its column matroid contains four pairwise disjoint bases.
For a set $A$ of column indices, let
$V=\langle v_i:i\in A\rangle$ and put $s=k-\dim V$.
If $s>0$, the subcode
$\{aG_R:a\in V^\perp\}$ has dimension $s$, and its support is
$\{i:v_i\notin V\}$. Therefore
\[
\#(\{1,\ldots,n\}\setminus A)
\ge\#\{i:v_i\notin V\}
\ge d_s(R)
\ge4(k-\dim V).
\]
For $s=0$ the same inequality is automatic.
By the matroid base-packing theorem \cite[Corollary~2.4]{BSY},
there exist four pairwise disjoint index sets $I_1,\ldots,I_4$,
each of size $k$, such that for each $j$ the columns indexed by $I_j$
form a basis of $\mathbb F_2^k$.

Expressing the residue and Fourier variables in this basis rewrites
\eqref{eq:hg-fourier-expansion} as
\begin{equation}
\frac{G_{A_4(C)}^{(g)}(Y)}{\Theta^n}
=d^{-k}\sum_{x=(x_1,\ldots,x_k)\in\mathcal H^k}
\chi_C(x)\prod_{i=1}^n b_{v_i\cdot x},
\label{eq:hg-column-contraction}
\end{equation}
where $\chi_C(x)\in\{1,-1\}$ and
$v_i\cdot x=\sum_{a=1}^k(v_i)_a x_a\in\mathcal H$.
For $1\le j\le4$, put
$F_j(x)=\prod_{i\in I_j}|b_{v_i\cdot x}|$.
By Lemma~\ref{lem:hg-quartic} and the disjointness of these index sets,
\[
\left|\prod_{i=1}^n b_{v_i\cdot x}\right|
\le\prod_{j=1}^4F_j(x).
\]
H\"older's inequality gives
\[
\sum_{x\in\mathcal H^k}\prod_{j=1}^4F_j(x)
\le\prod_{j=1}^4
\left(\sum_{x\in\mathcal H^k}F_j(x)^4\right)^{1/4}.
\]
For each $j$, the map
$x\mapsto(v_i\cdot x)_{i\in I_j}$ is a bijection of $\mathcal H^k$.
Consequently,
\[
\sum_{x\in\mathcal H^k}F_j(x)^4
=\left(\sum_{y\in\mathcal H}|b_y|^4\right)^k=d^k.
\]
Thus the absolute value of the sum in
\eqref{eq:hg-column-contraction} is at most $d^k$,
which proves \eqref{eq:hg-generalized-weight-bound}.
\end{pf}

\begin{ex}
\normalfont
\label{ex:hg-rm-one}
Let $m\ge5$ and consider the binary first-order Reed--Muller code
\[
\begin{aligned}
R_m=\operatorname{RM}(1,m)
=\left\{\left(a_0+\sum_{i=1}^m a_i x_i\right)_{x\in\F_2^m}
:a_0,a_1,\ldots,a_m\in\F_2\right\}\subseteq\F_2^{2^m}.
\end{aligned}
\]
It has dimension $k=m+1$, and its generalized Hamming weights are
\[
d_s(R_m)=
\begin{cases}
2^m-2^{m-s},&1\le s\le m,\\
2^m,&s=m+1.
\end{cases}
\]
Indeed, the common zero set of $s$ independent affine functions
is either empty or has $2^{m-s}$ elements. For $m\ge5$,
these values satisfy $d_s(R_m)\ge4s$ for all $s$.
For $m=5$, the weight hierarchy is
\[
(d_1,\ldots,d_6)=(16,24,28,30,31,32).
\]
The condition need not hold for smaller $m$; for instance,
$d_4(\operatorname{RM}(1,4))=15<16$.
Consequently, by Theorem~\ref{thm:hg-generalized-weights}, every
self-dual quaternary code $C$ with $\operatorname{Res}(C)=R_m$
satisfies, for all $g\ge1$ and $Y\in\operatorname{Sym}_g^+(\mathbb R)$,
\[
G_{A_4(C)}^{(g)}(Y)\le\Theta_g(Y)^{2^m}.
\]
{The quaternary code family
\(
C_{m,1}
=\operatorname{RM}(1,m)+2\operatorname{RM}(m-2,m)
\)
was constructed by Bonnecaze et al.
\cite[Sec.~IV-B, Proposition~7]{BSBM1997}.
These codes are Type~II self-dual and have
residue code $\operatorname{RM}(1,m)$.
Thus, for $m\ge5$, they provide an explicit
family to which Theorem~\ref{thm:hg-generalized-weights}
applies.
}
\end{ex}

The next lemma changes the code representation while preserving the
isometry class of its lattice. All vectors in its proof are row vectors.

\begin{lem}
\label{lem:hg-weight-four-reduction}
Let $C\subseteq\mathbb Z_4^n$ be self-dual and
$k=\dim\operatorname{Res}(C)$. If its residue contains a word of weight
$4$, then there is a self-dual code $C'\subseteq\mathbb Z_4^n$ such that
\[
A_4(C')\cong A_4(C),\
\dim\operatorname{Res}(C')\le k-1.
\]
\end{lem}

\begin{pf}
After a coordinate permutation, take
$r=(1,1,1,1,0,\ldots,0)$ in $R=\operatorname{Res}(C)$.
Choose a lift $c$ of $r$. Changing coordinate signs, an isometry of the
lattice, allows an integer representative
$c=(1,1,1,1,2h_5,\ldots,2h_n)$.
Let $L=A_4(C)$ and
$Q=\operatorname{diag}(H/2,I_{n-4})$, with $H$ as in
\eqref{eq:hg-hadamard}.

If $z/2\in L$, then $z\bmod2\in R$ and $R\subseteq R^\perp$ gives
$z_1+z_2+z_3+z_4\equiv0\pmod2$.
Since all entries of $H$ are odd, $zQ\in\mathbb Z^n$.
Thus $LQ\subseteq\tfrac12\mathbb Z^n$.
For $1\le i\le4$, the vector $2e_iQ^T$ is a row of $H$, extended by
zeros. Its reduction modulo $2$ is $r$, so
$2(2e_iQ^T)\bmod4=2r\in C$ and $2e_iQ^T\in L$.
For $i>4$, $2e_iQ^T=2e_i\in L$. Hence
\[
2\mathbb Z^n\subseteq LQ\subseteq\tfrac12\mathbb Z^n.
\]
Define $C'=(2LQ)\bmod4$. These inclusions imply $LQ=A_4(C')$.
Since $LQ$ is integral and unimodular, $C'$ is self-dual:
integrality gives self-orthogonality modulo $4$, and covolume $1$ gives
$\#C'=2^n$.

Extend $r$ to a basis of $R$ and choose corresponding lifts in $C$.
The integer lattice $2L$ is generated by these lifts, by $2t$ for
$t\in R^\perp$, and by $4\mathbb Z^n$.
The selected lift satisfies
$(1,1,1,1)H/2=(2,0,0,0)$, so $cQ$ is even.
For $t\in R^\perp$, the first four coordinates of $(2t)Q$ are
$(t_1,t_2,t_3,t_4)H$, which are even because $t\cdot r=0$.
The other coordinates are also even, as are all coordinates of
$4\mathbb Z^nQ$.
Only the remaining $k-1$ lifts can therefore contribute to the residue
of $C'$. This proves the dimension bound.
\end{pf}

\begin{thm}
\label{thm:hg-residue-rank-four}
Let $C\subseteq\mathbb Z_4^n$ be a self-dual code satisfying
$\dim_{\mathbb F_2}\operatorname{Res}(C)\le4$.
Then, for every $g\ge1$ and every
$Y\in\operatorname{Sym}_g^+(\mathbb R)$,
\begin{equation}
G_{A_4(C)}^{(g)}(Y)\le\Theta_g(Y)^n.
\label{eq:hg-rank-four-bound}
\end{equation}
\end{thm}

\begin{pf}
We argue by induction on $k=\dim R$, where $R=\operatorname{Res}(C)$.
For $k=0$, the lattice is $\mathbb Z^n$.
If $R$ contains a word of weight $4$, apply
Lemma~\ref{lem:hg-weight-four-reduction} and the induction hypothesis.
Gaussian mass is invariant under lattice isometries.
We may therefore assume that every nonzero word of $R$ has weight at
least $8$.

For an $s$-dimensional subcode $U\le R$, each coordinate in its support
is nonzero in exactly $2^{s-1}$ words of $U$. Hence
\begin{equation}
\#\operatorname{supp}(U)
=2^{1-s}\sum_{0\ne u\in U}\operatorname{wt}(u)
\ge8(2-2^{1-s}).
\label{eq:hg-subcode-average}
\end{equation}
In particular, $d_1(R)\ge8$, $d_2(R)\ge12$, and $d_3(R)\ge14$.
For $k\le3$, Theorem~\ref{thm:hg-generalized-weights} applies directly.
For $k=4$, it also applies if
$N=\#\operatorname{supp}(R)\ge16$.

It remains to consider $k=4$ and $N\le15$.
Equation~\eqref{eq:hg-subcode-average} gives $N\ge15$, so $N=15$ and
every nonzero residue word has weight $8$.
Coordinates outside the support of $R$ split off copies of $\mathbb Z$:
if the $i$th residue coordinate is zero, then $e_i\in R^\perp$,
$2e_i\in C$, and $e_i\in A_4(C)$; all lattice vectors have integral
$i$th coordinate. We can therefore restrict to these fifteen coordinates
and multiply the final inequality by $\Theta^{n-15}$.

Choose a binary generator matrix for the restricted residue and let
$m_v$ denote the multiplicity of its column $v\in\mathbb F_2^4$.
For $v\ne0$, Walsh inversion gives
\[
m_v=-\frac18\sum_{a\in\mathbb F_2^4}
(-1)^{a\cdot v}\operatorname{wt}(aG_R)=1.
\]
Thus the residue is the binary simplex code $[15,4,8]$.

We next show that all self-dual lifts of this residue are equivalent by
coordinate sign changes. Choose residue generators $r_1,\ldots,r_4$ and
lifts $c_a=r_a+2h_a$, and put
$M_{ab}=r_b\cdot h_a$ in $\mathbb F_2$.
Distinct generators have integer inner product $4$, so
self-orthogonality gives $M_{ab}=M_{ba}$.
The matrix $M$ determines the code: it determines each $h_a$ modulo
$R^\perp$, and the torsion is $2R^\perp$.
Changing the sign of a coordinate of column type $v$ adds $vv^T$ to
$M$. All nonzero types occur, and the matrices $vv^T$ span the space of
symmetric $4\times4$ binary matrices. Indeed, $v=e_i$ gives the diagonal
matrices, while $v=e_i+e_j$, together with the diagonals, gives each
symmetric off-diagonal matrix.
Consequently every lift is sign-equivalent to
\[
C_0=\langle r_1,r_2,r_3,r_4,2R^\perp\rangle_{\mathbb Z_4}.
\]
This code is self-dual: the integer residue generators have weights $8$
and pairwise inner products $4$, the torsion is orthogonal to them, and
$\#C_0=2^{15}$.

Order the fifteen coordinates by the nonzero column types
$v=1,\ldots,15$, where the four entries of $v$ are its binary digits,
starting with the least significant digit.
Retain the first three coordinates and set
\[
B_1=\{4,5,6,7\},\
B_2=\{8,9,10,11\},\
B_3=\{12,13,14,15\}.
\]
Apply the orthogonal transformation
\[
Q_{15}=\operatorname{diag}(I_3,H/2,H/2,H/2).
\]
Every residue word has even weight on each block $B_j$, so
$2A_4(C_0)Q_{15}\subseteq\mathbb Z^{15}$.
Also, the indicator of each $B_j$ belongs to $R^\perp$.
The inverse images of the new frame vectors $2e_i$ consequently belong
to $A_4(C_0)$, exactly as in
Lemma~\ref{lem:hg-weight-four-reduction}.
Thus $A_4(C_0)Q_{15}=A_4(C_0')$ for a self-dual code $C_0'$.

The transformed lifts of $r_3$ and $r_4$ are even, because
\[
\operatorname{supp}(r_3)=B_1\cup B_3,\
\operatorname{supp}(r_4)=B_2\cup B_3,
\]
and $(1,1,1,1)H/2=(2,0,0,0)$.
For $t\in R^\perp$, write
$\tau_j=\sum_{i\in B_j}t_i$ in $\mathbb F_2$.
Orthogonality to $r_3$ and $r_4$ gives
$\tau_1=\tau_2=\tau_3$.
On $B_j$, the residue of $(2t)Q_{15}$ is
$\tau_j(1,1,1,1)$; on the first three coordinates it is zero.
All torsion generators therefore contribute at most one residue
direction, while the transformed lifts of $r_1$ and $r_2$ contribute at
most two. The images of $4\mathbb Z^{15}$ are even.
It follows that $\dim\operatorname{Res}(C_0')\le3$.

The already established case $k\le3$, together with lattice isometry and
the split-off integer coordinates, proves
\eqref{eq:hg-rank-four-bound}.
\end{pf}
}
\section{Conclusion}\label{conclusion}

In this paper we have explored inequalities on the Gaussian mass of unimodular lattices of the type
\[
\forall t\in(0,\infty),\ G_L(t)\le G_{\Z^n}(t).
\]

We have shown that they hold exactly for extremal unimodular lattices of both types I and II and on average for
\begin{itemize}
\item
all Type II lattices in a given dimension,
\item lattices obtained by Construction A from free self-dual $\Z_4$-codes,
\item lattices obtained by Construction A  from
 self-dual $\Z_6$-codes.
 \end{itemize}
  Similar results have been derived for higher genus theta series, in particular the direct analogue of \cite{SDK} for the binary Construction A.
 It seems very hard to prove the $\Z_4$ analogue of \cite{SDK} (even for $g=1$!) with the current technology. This is the main problem left open by this research.\\
\ \

\noindent
{\bf Acknowledgement:} The last author is grateful to Professor Kominers for suggesting to look at higher genus theta series, and to Professor Bollauf for helpful discussions.\\
\ \

\noindent
{\bf Declaration of AI use:} The authors used chatGPT (Open AI) for  several technical proofs. All the AI-assisted content was reviewed, verified, and revised by the authors, who take full responsibility for the results, the arguments, and the final form of the article.

\end{document}